\documentclass[draft]{agujournal2019}
\usepackage{url} 
\usepackage{lineno}
\usepackage[inline]{trackchanges} 
\usepackage{soul}

\usepackage{booktabs}
\usepackage{colortbl}
\usepackage{amsmath}
\usepackage{MnSymbol}

\drafttrue

\journalname{JGR: Oceans}

\begin{document}

%
%


\title{Quantifying Bore-bore Capture on Natural Beaches}

%
%




\authors{C. E. Stringari\affil{1}, H. E. Power\affil{1}}


\affiliation{1}{University of Newcastle, School of Environmental and Life Sciences, Newcastle, Australia}




\correspondingauthor{C. E. Stringari}{Caio.EadiStringa@uon.edu.au}




\begin{keypoints}
\item There is a high probability ($\approx$40\%) of bore-bore capture occurring in the surf or swash zones.
\item Bore-bore capture occurs under infragravity wave energy dominance only $\approx$50\% of the time.
\item The majority ($>$97\%) of extreme shoreline maxima were directly driven by bore-bore capture.
\end{keypoints}

%
%

%
%


\begin{abstract}
Bore-bore capture occurs when a faster moving bore captures a slower moving bore whilst both are propagating shoreward in the surf or swash zones. This phenomenon occurs frequently on natural beaches, but has not yet been quantified in the literature. Novel application of wave tracking methods allowed for investigation of this phenomenon at seven sandy, micro-tidal, wave-dominated Australian beaches. The results showed that, for the locations where beach slope and environmental conditions allowed for bore-bore capture to occur, there was a high probability ($\approx$40\%) of one bore capturing another bore in the surf or swash zones. The landward-most 10\% of the nearshore region (i.e., the time-varying surf-swash extent) was found to be the most likely location for a bore-bore capture event. Amplitude and frequency dispersion and the interaction between bores and infragravity waves are (indirectly) shown to to be equally important for driving bore-bore capture. Capture events infrequently led to extreme horizontal shoreline maxima (approximately 20\% of the cases), however, most extreme shoreline maxima were directly driven by bore-bore capture events ($>$97\% of the cases). For the analysed data, there was a direct relation between the probability of bore-bore capture driving extreme shoreline maxima and the Iribarren Number and beach morphodynamic state: the steeper or more reflective the beach, the higher the probability of a bore-bore capture event causing an extreme shoreline maxima event. Such correlation has direct importance for the future development of predictive runup models, which currently do not account for this phenomenon.
\end{abstract}

\section*{Plain Language Summary}
An observer looking at waves on a beach will notice that sometimes one faster-moving broken wave will overtake and capture another broken wave. This phenomenon, known as bore-bore capture, occurs frequently on natural beaches and this study presents a precise quantification of this phenomenon for the first time. Bore-bore capture is important for several aspects of the beach, for example, moving great amounts of sand and for general beach goer’s safety, as bore-bore capture may cause higher-than-normal water excursions. Here we use video data collected at seven Australian beaches and modern artificial intelligence techniques to track the evolution of the waves as they travel across the beach and precisely quantify where and how many broken waves were captured by other broken waves. Our results showed that approximately 40\% of waves were captured by another wave and that the great majority (97\%) of extreme water excursions on the beach were directly generated by bore-bore capture events. These results are important because they show that bore-bore capture needs to be better accounted for in the tools that coastal engineers and managers use to manage the coast and ensure the safety of beach-goers.

%
%

\section{Introduction}\label{Chap_5_Sec_1}

Wave or bore merging \cite{Tissier2015}, bore-bore capture \cite{Garcia-Medina2017}, wave focusing \cite{Alsina2018b}, swash overtake \cite{Chardon-Maldonado2016}, or wave overrunning \cite{Power2015}, hereafter referred to as bore-bore capture, all describe the process of one bore (i.e., a  broken wave) capturing another bore in the surf or swash zones (hereafter collectively referred to as the nearshore \cite{komar1976beach}). This topic has received increased research interest in the past few years due to the implications it may have for the surf-swash boundary, especially for sediment transport dynamics \cite{Alsina2018b} and extreme runup heights \cite{Garcia-Medina2017}. Due to the dispersive characteristics of water waves in shallow water, changes in the local depth ($h$) directly imply changes in the wave speed ($c$) \cite{svendsen2006}. For instance, laboratory data has recently shown that shoreward propagating infragravity waves (with frequencies between 0.004 and 0.04Hz; hereafter IG waves) alter the water depth in which sea-swell waves (with frequencies between 0.04 and 1.0Hz; hereafter SW waves) are propagating, leading to bore-bore captures \cite{Tissier2015, VanDongeren2007}. Although this phenomenon occurs frequently on natural beaches \cite{Atkinson2017, Bradshaw1982, Moura2018, Guedes2013, Huntley1976, Senechal2001}, a direct quantification has not been attempted to date.

Frequency and amplitude dispersion, along with wave-wave interactions, cause changes in the SW wave speed ($c_{sw}$), with frequency dispersion causing shorter waves to propagate slower than longer waves, while amplitude dispersion causes larger waves to propagate faster than smaller waves \cite{svendsen2006}. Interactions between IG and SW waves modulate the water depth in which SW waves are propagating and also alter SW wave orbital velocities \cite{Tissier2015}, therefore, affecting $c_{sw}$. To add more complexity, bound \cite{LonguetHiggins1964}, break-point \cite{Symonds1982}, and edge waves \cite{Bowen1978} may coexist in the surf zone \cite{Bertin2018b}. These different types of IG waves can propagate, shoal, and dissipate energy all while interacting with SW waves and other nearshore waves \cite{Battjes2004, VanDongeren2007}. Finally, surf zone currents interact with the incoming SW and IG waves, also modulating $c_{sw}$ \cite{Svendsen2003, Nam2009, MacMahan2010, Almar2016}. All these phenomena must be accounted for to enable a precise physical description of bore-bore capture. Due to these complexities, this work will focus on describing the natural inter- and intra-beach variability of bore-bore capture rather than analysing the underlying hydrodynamics in depth.

This paper uses the computer vision, machine learning, and data-mining techniques from \citeA{Stringari2019} to directly detect bore-bore capture in time-space (timestack) images \cite{Aagaard1989}. Novel, non-stationary analysis (wavelet-based) is used to test the hypothesis that bores propagating on the crest of IG waves are more likely to capture other bores in the surf zone. Additionally, extreme event analyses are used to evaluate whether bore-bore captures lead to extreme horizontal shoreline maxima. This paper is organised as follows. Section \ref{Chap_5_Sec_2} outlines the data being used throughout the paper, presents the novel method for bore-bore capture detection, describes how non-stationary surf and swash zone extents are measured, and describes how IG wave phase and energy data are calculated. Section \ref{Chap_5_Sec_3} presents results as statistical descriptions of bore-bore capture and analyses of the correlation between extreme horizontal shoreline maxima and bore-bore capture. Finally, Section \ref{Chap_5_Sec_4} presents a discussion, and Section \ref{Chap_5_Sec_5} provides a conclusion.

\section{Materials and Methods}\label{Chap_5_Sec_2}

\subsection{Data Collection and Pre-processing}\label{Chap_5_Sec_2_Sub_1}

Video imagery, pressure transducer (PT) and topographic survey data were collected at seven sandy, micro-tidal, wave-dominated Australian beaches. The experiments were conducted at One Mile Beach (Forster, New South Wales (NSW), hereafter  OMB), Werri Beach (Gerringong, NSW, hereafter WB), Moreton Island Eastern Beach (Queensland, hereafter MI), Frazer Park Beach (Frazer Park, NSW, hereafter FB), Seven Mile Beach (Gerroa, NSW, hereafter SMB), Nobbys Beach (Newcastle, NSW, hereafter NB), and Elizabeth Beach (Pacific Palms, NSW, hereafter EB). These beaches were chosen because they cover a wide range of morphodynamic states \cite{Wright1984} and have elevated headlands or sand dunes that allow for video collection. Based on visual classification of Timex images and data from \citeA{short1999beaches, short2007beaches}, SMB was identified as representing the dissipative (D) state, MI the alongshore bar and trough (LBT) state, FB and NB the transverse bar and rip state (TBR) state, OMB and WB the low tide terrace (LTT) state during mid and low tides, and EB the LTT state during high tide. 

\begin{figure}[htp]
	\centering
	\includegraphics[width=0.95\textwidth]{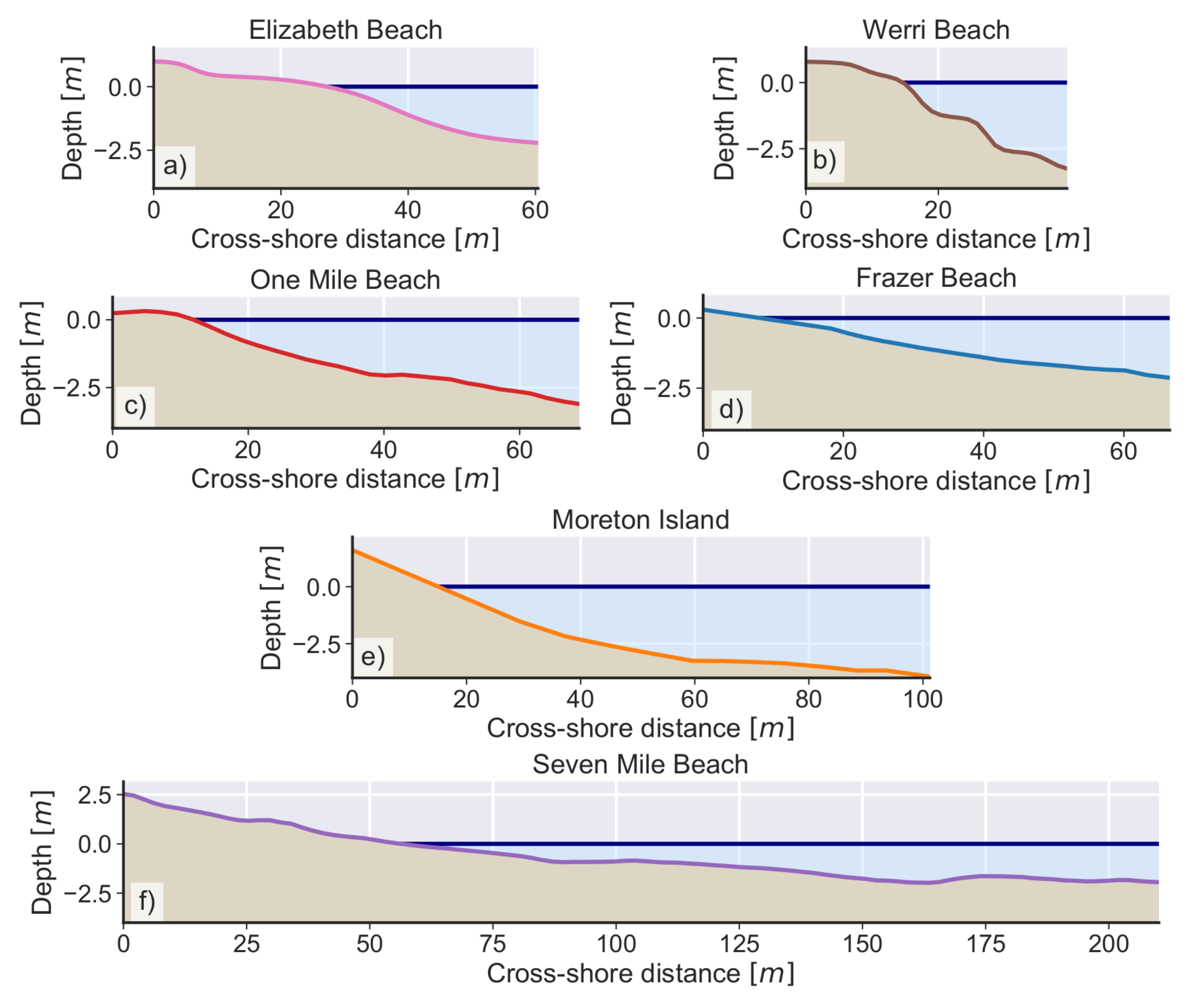}
	\caption[Representative beach profiles for Elizabeth Beach, Werri Beach, One Mile Beach, Frazer Beach, Moreton Island, and Seven Mile Beach.]{Representative beach profiles for: a) Elizabeth Beach, b) Werri Beach, c) One Mile Beach, d) Frazer Beach, e) Moreton Island, and f) Seven Mile Beach. The thick blue line shows the mean water level calculated based on the offshore-most PT over the duration of each deployment. All subplots have the same aspect ratio (1/4).}\label{Chap5_Fig_01}
\end{figure}

For each experiment, the nearshore region was videoed from a headland or elevated dune using a consumer grade Sony video camera (Sony HDR-XR200 for the 2014 experiments and Sony HDR CX240 for the remainder of the experiments). PTs were deployed in a cross-shore transect recording at a minimum sample frequency of 8Hz and a beach survey (profile and ground control points) was acquired using a total station. The NB dataset was collected during storm conditions and therefore, due to safety reasons, no PTs were deployed nor was a sub-aqueous beach profile surveyed during this deployment. This deployment was designed specifically to study the effect of storm conditions on bore-bore capture (See Section \ref{Chap_5_Sec_4_Sub_3}). Figure \ref{Chap5_Fig_01} shows representative profiles for each location and Table \ref{Chap5_Tab_01} shows summarised nearshore and offshore data for each experiment.

\begin{sidewaystable}
	\caption[Data record summary for each experiment.]{Data record for each experiment. nPTs is the number of PTs in each deployment.  $H_{m_0}$  and  $T_{m_{01}}$ are the observed surf zone significant wave height and period at PT closest to the middle of the surf zone (see Section \ref{Chap_5_Sec_2_Sub_2} for definitions). $H_{m_{0\infty}}$, $T_{m_{01\infty}}$ and $D_m$ are the offshore significant wave height, peak period, and peak direction respectively. B. Or. is the beach orientation relative to the geographic north. $\beta$ is the representative beach slope calculated in the swash and inner surf zones, and M. State is the visually observed beach morphodynamic state \cite{Wright1984}. $\Omega_{\infty}=\frac{H_{m_{0\infty}}}{T_{m_{01\infty}} W_s}$, where $W_s$ is the sediment fall velocity and $\xi_\infty=\frac{\tan{\beta}}{\sqrt{H_{m_{0\infty}} L_\infty}}$, where $L_\infty=\frac{g}{2\pi}T_{m_{01\infty}}^2$.}\label{Chap5_Tab_01}
	\centering
	\begin{tabular}{llllllllllllll}
		\toprule
		Location & Date & nPTs  & $H_{m_0}$ & $T_{m_{01}}$  & $H_{m_{0\infty}}$ & $T_{m_{01\infty}}$ & $D_m$ & B. Or. & $\beta$ & $W_s$ & $M. State$ & $\Omega_\infty$ &  $\xi_\infty$  \\
		\midrule
		WB~  & 16/08/2014~ & 8  & 0.95 & 9.5  & 1.55 & 12.50 & ESE~ & NE & 0.11 & 0.07 & LTT~     & 1.90 & 4.49  \\
		OMB~ & 7/08/2014~  & 9  & 0.75 & 10.5 & 1.63 & 9.00  & S~   & SE & 0.05 & 0.05 & TBR/LTT~ & 3.40 & 1.33  \\
		MI~  & 20/12/2016~ & 14 & 0.82 & 8.0  & 1.46 & 10.50 & SE~  & E  & 0.07 & 0.04 & LBT~     & 3.48 & 2.35  \\
		SMB~ & 14/06/2018~ & 30 & 1.05 & 10.5 & 1.25 & 11.50 & SE~  & S  & 0.03 & 0.03 & D~       & 3.77 & 1.21  \\
		NB~  & 02/11/2017~ & -~ & -~   & -~   & 3.02 & 10.40 & SE~  & E  & 0.07 & 0.05 & TBR~     & 6.15 & 1.67  \\
		FB~  & 24/04/2018~ & 11 & 0.43 & 6.5  & 1.75 & 8.75  & SE~  & SE & 0.04 & 0.03 & TBR~     & 6.17 & 0.91  \\
		EB   & 13/05/2019  & 12 & 0.85 & 11.7 & 1.05 & 13.00 & SE~  & NE & 0.27 & 0.05 & LTT~     & 1.70 & 4.74 \\
		\bottomrule
	\end{tabular}            
\end{sidewaystable}

Video data were processed following the standard ARGUS methodology \cite{Holland1997, Hoonhout2015}. From the raw video record (25Hz), frames were extracted at 10Hz and projected into metric coordinates using the surveyed ground control points. The image coordinate referential was rotated and translated so that the x-coordinate was oriented shoreward and fully aligned with the surveyed profile and sensor transect. From this, a timestack was extracted at the same location as the PT transect and beach profile. For NB, due to the lack of a PT transect, the timestack line was extracted at the most representative location. Individual waves were tracked and optimal wave paths and their respective confidence intervals were obtained for each timestack  as per \citeA{Stringari2019}. For each location, one hour of data was processed in five-minute batches, totalling 2736 individual waves being tracked. This population size results in an error $<$1$\pm$0.5\%  at the 99\% confidence level \cite{Freedman1998}. To ensure that all waves were tracked correctly, all tracking errors were corrected manually using the QGIS interface provide alongside the tracking algorithm, which was found to be considerably time-consuming and therefore limited the size of the final dataset (see  \citeA{Stringari2019} for further discussion on this issue).

\subsection{Bore-bore Capture, Surf Zone Limits and Shoreline Detection}\label{Chap_5_Sec_2_Sub_2}

\begin{figure}[htp]
	\centering
	\includegraphics[width=0.95\textwidth]{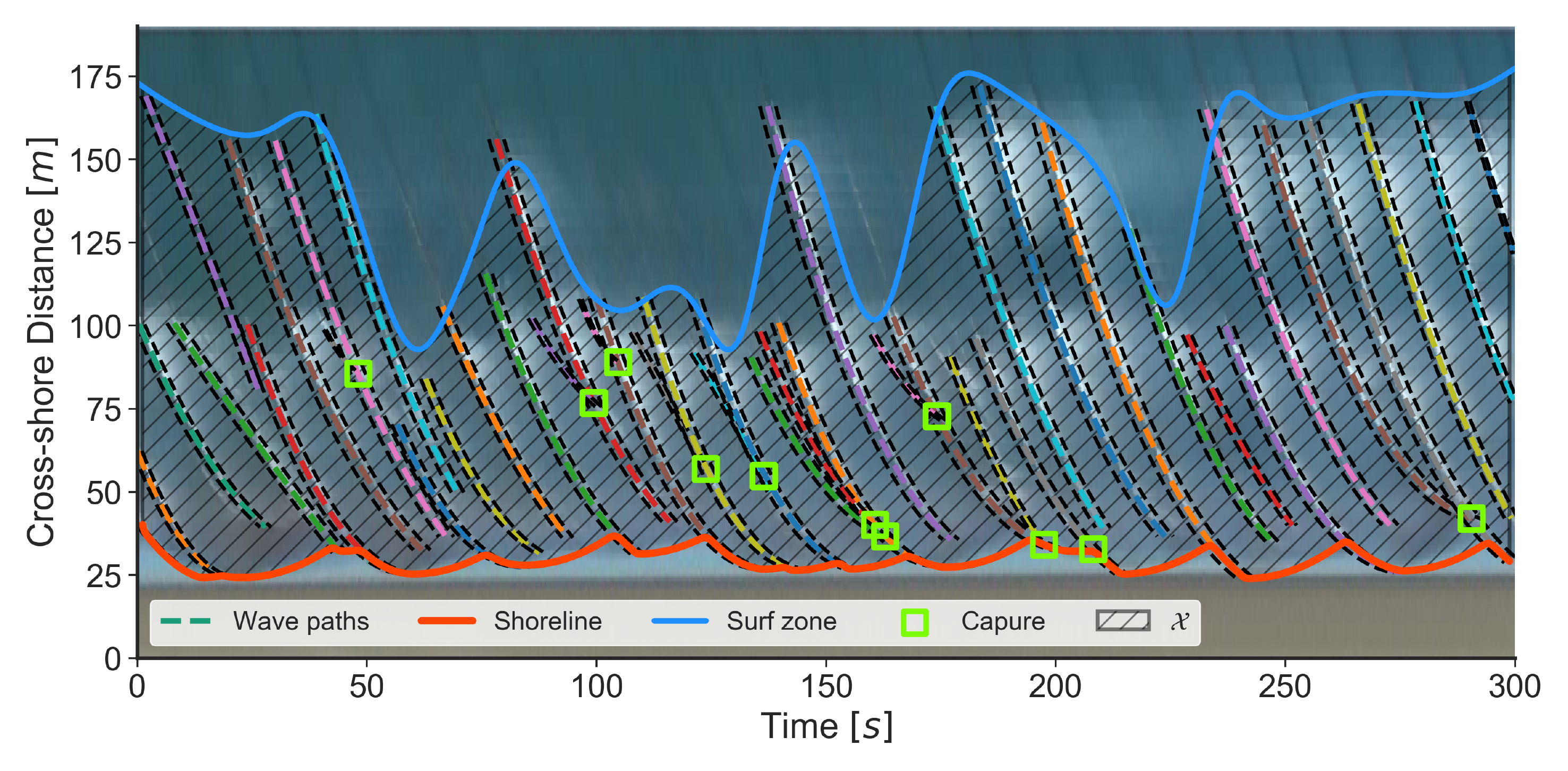}
	\caption[Example of the bore-bore capture detection algorithm for five minutes of data collected at Seven Mile Beach.]{Example of the bore-bore capture detection algorithm for five minutes of data collected at SMB. The dashed coloured lines show the optimal wave paths and the dashed black lines the confidence intervals for each wave path. The light green square markers show bore-bore capture occurrences. The continuous red and blue lines show the seaward limit of the surf zone and landward limit of the swash zone, respectively. The hatched area is a graphical representation of the non-stationary nearshore width (see Section \ref{Chap_5_Sec_2_Sub_1}).}\label{Chap5_Fig_02}
\end{figure}

Bore-bore capture events were detected directly from the tracked wave paths. A capture event was defined as the intersection of two adjacent wave paths within the combined wave paths’ confidence intervals (green markers in Figure \ref{Chap5_Fig_02}). This definition is implemented by solving systems of equations of the form:

\begin{align}
x_1&=x_2\nonumber \\
a_1t^2_1+b_1t+c_1&=	a_2t^2_2+b_2t+c_2
\end{align}\label{Chap_5_Eq_1}

\noindent in which the subscripts $\{1,2\}$ represent two different wave paths, $x$ is the cross-shore position, $t$ is time, and $\{a, b, c\}$  are coefficients to be learnt via the optimisation procedure. The root of these systems of equations (if any) represent the intersections of two adjacent wave paths. The implementation was done in python using \textit{SciPy}’s built-in linear algebra solvers \cite{Jones2001}. These intersections were then overlaid into the original timestacks and visually checked to ensure that all events were detected. No instances of wrongly identified or missed capture events were observed. Note that bore-bore capture events were detected prior to obtaining the downrush motion detailed below, therefore, no instances of uprush capturing a downrush were included in the data.

The time-varying surf zone limit and shoreline were also computed directly from the wave paths (continuous blue and red lines in Figure \ref{Chap5_Fig_02} respectively). The seaward limit of surf zone was calculated as follows. For each timestamp, the outer-most edge of the tracked waves was identified and interpolated to continuous time vector using a Gaussian radial basis function (RBF, \cite{Iske2004}) of the form:

\begin{equation}
\Phi=e^{-(\nu(r))^2}
\end{equation}

\noindent in which $r=t-t_1$ and $\nu$ is the inverse of the critical radius, or shape parameter. This algorithm can be seen as an automated alternative to the method developed by \citeA{DeMoura2017} and is transferable to other applications. The time-varying landward limit of the swash zone was calculated in three steps. First, the uprush was obtained directly from the wave paths. Second, the downrush was predicted using a ballistic equation \cite{shen1963, Brocchini2008}:

\begin{equation}
Dr(t)=\frac{1}{2}\left( \frac{1}{\sqrt{1+\beta^2}} \right)t^2 + Up_{0} 
\end{equation}

\noindent where $Dr(t)$ is the time-dependent downrush position and $Up_0$ is the initial position of the motion, here assumed to be the maximum uprush of an incoming wave path. Third, the two swash motions were interpolated to a continuous time vector using the same RBF function as before and smoothed using a moving average model with a Gaussian kernel. Although relatively simplistic, this method produced good results with mean absolute percentage errors (MAPE) of less than 2\% when compared to manually digitised shorelines (see Table \ref{Chap5_Tab_02}).

\begin{table}
	\caption[Errors in the time-varying shoreline detection.]{Errors in the time-varying shoreline detection. The comparison was done using 15 minutes of manually digitised shoreline data for each location. MAE is the mean absolute error, RMSE is the root mean squared error, MAPE is the mean absolute percentage error, and corr. coef. is the Pearson’s product-moment correlation coefficient. The cross-shore horizontal resolution is 0.1m for all timestacks. $^{*}$The shoreline variation was not tracked at Elizabeth beach due to the absence of bore-bore captures (see Section \ref{Chap_5_Sec_3_Sub_1}).}\label{Chap5_Tab_02}
	\centering
	\begin{tabular}{lllll}
		\toprule \\
		Location         & MAE [m] & RMSE [m] & MAPE [\%]           & Corr. Coef [-]  \\
		\midrule \\
		Frazer Beach     & 0.214   & 0.211    & 0.84                & 0.99            \\
		Moreton Island   & 0.012   & 0.358    & 0.05                & 0.9             \\
		Nobbys Beach     & 0.162   & 0.051    & 0.19                & 0.99            \\
		One Mile Beach   & 0.09    & 0.139    & 0.41 & 0.94            \\
		Seven Mile Beach & 0.392   & 0.672    & 1.64 & 0.98            \\
		Werri Beach      & 0.145   & 0.101    & 0.49                & 0.97 \\
		Elizabeth Beach$^{*}$ & -   & -    & -  & - \\
		\bottomrule \\           
	\end{tabular}
\end{table}

\subsection{Quantifying Infragravity Wave Influence on Bore-bore Capture}\label{Chap_5_Sec_2_Sub_3}

The continuous wavelet transform  \cite{Torrence1998} was used to quantify the time-dependent IG amplitude, phase and energy. The transform was directly applied to a 7-minute timeseries of the surface elevation record ($\eta$) centred at each 5-minute data run for the PT nearest to the middle of the surf zone (see Figure \ref{Chap5_Fig_02} for the definition of the surf zone). Varying the PT location did not change the results significantly (not shown). Further, in the case of a progressive free IG wave that maintains its height through the surf zone, this approach is valid given that $c_{sw}$ and $c_{ig}$ are equivalent in shallow water under weakly dispersive conditions. This means that an SW wave on the crest of IG wave stays on the IG crest while both propagate shoreward \cite{Tissier2015, VanDongeren2007}. Alternatively, in the case of an IG wave that decays while propagating throughout the surf zone \cite{Baldock2012}, the influence of the IG wave on bore-bore capture would, therefore, be even further diminished. 

The continuous wavelet transform $W_n(s)$ of a timeseries $X_n$ $(n={1,2,3...,N-1})$ is defined as the convolution of $X_n$ with a wavelet function  such that:

\begin{equation}
W_n(s) = \sum_{n'=0}^{N-1} X_{n'}\psi*\left(\frac{(n-n) dt}{s} \right)
\end{equation}

\noindent in which $dt$ is the sampling interval, $*$ denotes the complex conjugate operator, and  $s$ is a scale parameter. Scaling and translating $\psi$ in time allows for the decomposing of the signal in both time ($t$) and frequency ($f$) domains. In this work, the complex Mexican Hat wavelet was used as the mother wavelet, which is defined in non-dimensional time ($\tau$) as:

\begin{equation}
\psi(\tau)=\frac{2}{\sqrt{3}}\pi^{-\frac{1}{4}}\left(\sqrt{\pi}(1-\tau^2)e^{-\frac{1}{2}\tau^2}-\left(\sqrt{2}i\tau+\sqrt{\pi}\operatorname{erf}\left[\frac{i}{\sqrt{2}}t\right]\left(1-\tau^2\right)e^{-\frac{1}{2}\tau^2}\right)\right)
\end{equation}

\noindent where $\operatorname{erf}$ is the Gaussian error function \cite{Addison2002}. The implementation was done based on \citeA{Grinsted2004} with the only modification being the addition of the complex Mexican Hat wavelet.

The normalised IG wave amplitude was obtained by zeroing the contributions outside of the IG frequency band in the wavelet local spectrum ($W$) and applying the inverse wavelet transform (panel b) in Figure \ref{Chap5_Fig_03}, orange line). The time-dependent IG wave phase was obtained from the frequency-averaged complex part of wavelet transform. It was also useful to define time-dependent absolute and relative energy contributions in each frequency band (panels d) and e) in Figure \ref{Chap5_Fig_03}). These energy contributions are defined as:

\begin{equation}
E_{sw}(t) = \int_{f=0.04}^{f=1}|W|^2(t, f)df
\end{equation}

\begin{equation}
E_{ig}(t) = \int_{f=0.004}^{f=0.04}|W|^2(t, f)df
\end{equation}

\noindent All integrations were done using the trapezoidal rule. Comparing the time-dependent IG wave amplitude, energy, and phase to the bore-bore capture occurrences allowed for identification of whether a bore-bore capture occurred on the crest (positive phase) or trough (negative phase) of an IG wave and whether IG or SW energy was dominant in the surf zone during the given capture event. The results of this analysis are presented in Section \ref{Chap_5_Sec_3_Sub_3}.

\begin{figure}[htp]
	\centering 
	\includegraphics[width=0.99\textwidth]{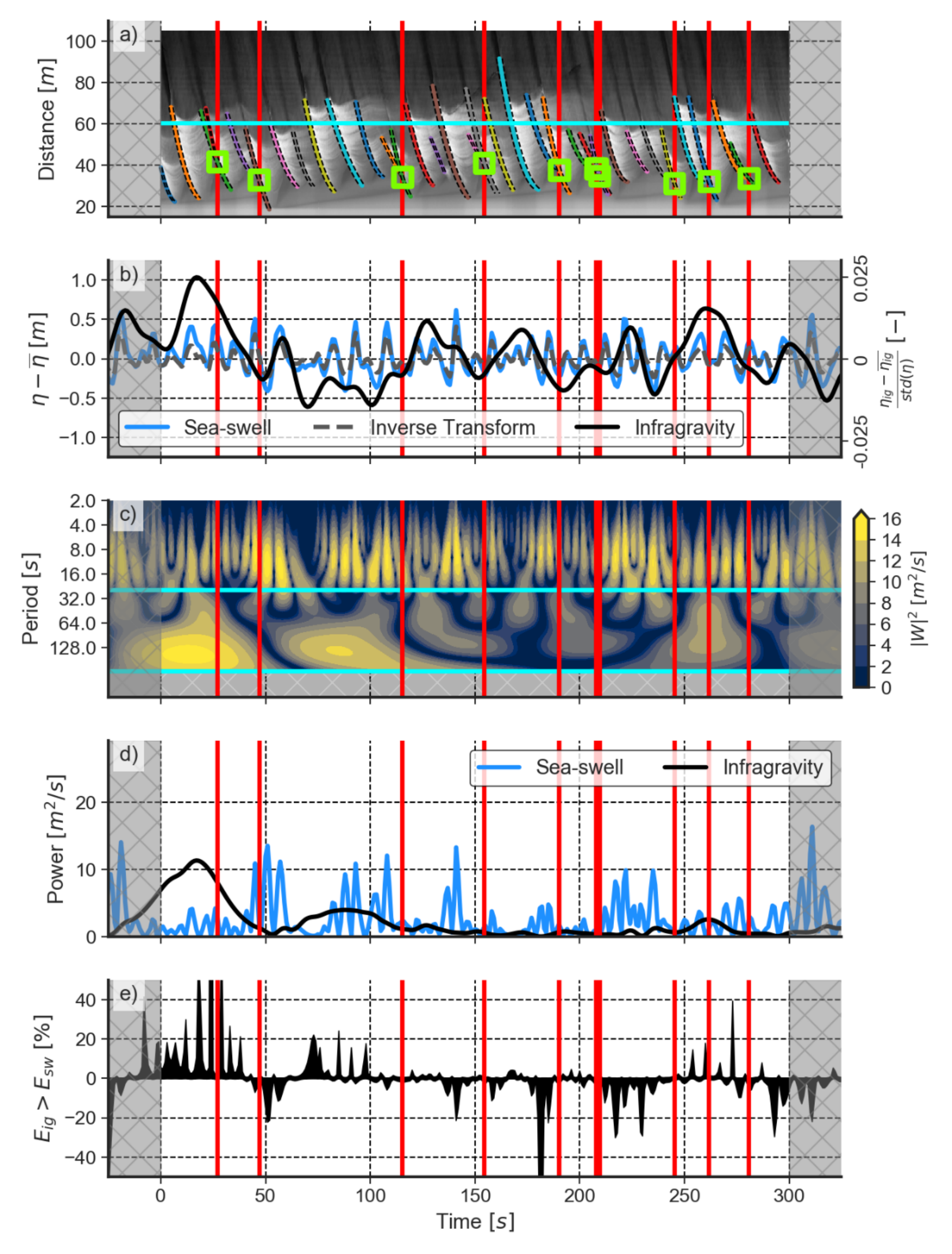}
	\caption{(Caption next page.)}\label{qb_by_depth}
\end{figure}
\addtocounter{figure}{-1}
\begin{figure}[t!]
	\caption[Example of the wavelet transform analysis applied to One Mile Beach data.]{Example of the wavelet transform analysis applied to One Mile Beach data. a) Timestack showing the tracked wave paths and bore-bore capture events (light green markers and red vertical lines). The cyan line shows the location of the PT timeseries used in b), c), d), and e). b) Surface elevation record in both SW (blue, left axis) and IG (orange, right axis) frequency bands. The dashed grey line in b) shows the inverse transform in the SW band. c) Wavelet transform power spectrum density ($W_n$). Lighter coloured regions show more energetic portions of the spectrum. The hatched areas show regions out-of-the-analysis-domain (left and right) and out-of-the-cone-of-influence regions. The light blue lines show the SW and infragravity bands limits. d) Integrated local power spectrum in the SW (blue) and  IG frequency bands (orange). e) Timeseries showing when $E_{ig}(t)$ is greater than $E_{sw}(t)$ ($\%$). The red lines in all panels show the time occurrence of bore-bore captures.}\label{Chap5_Fig_03}
\end{figure}

\section{Results}\label{Chap_5_Sec_3}

\subsection{Bore-bore Capture Probabilities}\label{Chap_5_Sec_3_Sub_1}

Bore-bore capture was observed on six of the seven analysed beaches with bore-bore capture not observed at Elizabeth Beach, likely due to the steep beach profile observed at this location. For the beaches where bore-bore capture was observed, any given broken wave had, on average, a 38\% probability of being captured by another broken wave while propagating in the surf or swash zones (see Table \ref{Chap5_Tab_03}). The average bore-bore capture return period was 24.0s, which is marginally outside the lower IG frequency cut-off. Even for the most gently sloping beach in the dataset (SMB), the average capture period (20.2s) was well below the IG frequency limit (25s). For the discussion of these results, see Section \ref{Chap_5_Sec_4}.

\begin{table}
	\caption{Averaged bore-bore capture statistics. In this analysis, the mean wave period was calculated dividing the number of tracked waves by the total length of the record, which is equivalent to the zero-crossings wave period ($T_{m_{02}}$) \cite{holthuijsen2010}. The statistics indicated with $^*$ did not include EB data.}\label{Chap5_Tab_03}
	\begin{tabular}{lllllllll}
		\toprule
		& FB    & MI    & NB    & OMB   & SMB   & WB    & EB  & All loc.  \\
		\midrule
		Number of waves          & 546   & 507   & 418   & 391   & 476   & 415   & 401 & 2736         \\
		Number of merges         & 203   & 166   & 218   & 111   & 176   & 101   & 0   & 975          \\
		Mean wave period         & 6.69  & 7.08  & 8.33  & 9.12  & 7.4   & 8.61  & -   & 7.87$^*$        \\
		Mean capture period      & 17.73 & 21.69 & 16.51 & 32.43 & 20.45 & 35.64 & -   & 24.08$^*$       \\
		Capture probability & 0.41  & 0.34  & 0.55  & 0.31  & 0.4   & 0.25  & -   & 0.38$^*$ \\       
	\end{tabular}
	\centering
\end{table}

\subsection{Most Likely Capture Location}\label{Chap_5_Sec_3_Sub_2}

\begin{sidewaysfigure}[htp]
	\centering 
	\includegraphics[width=0.99\textwidth]{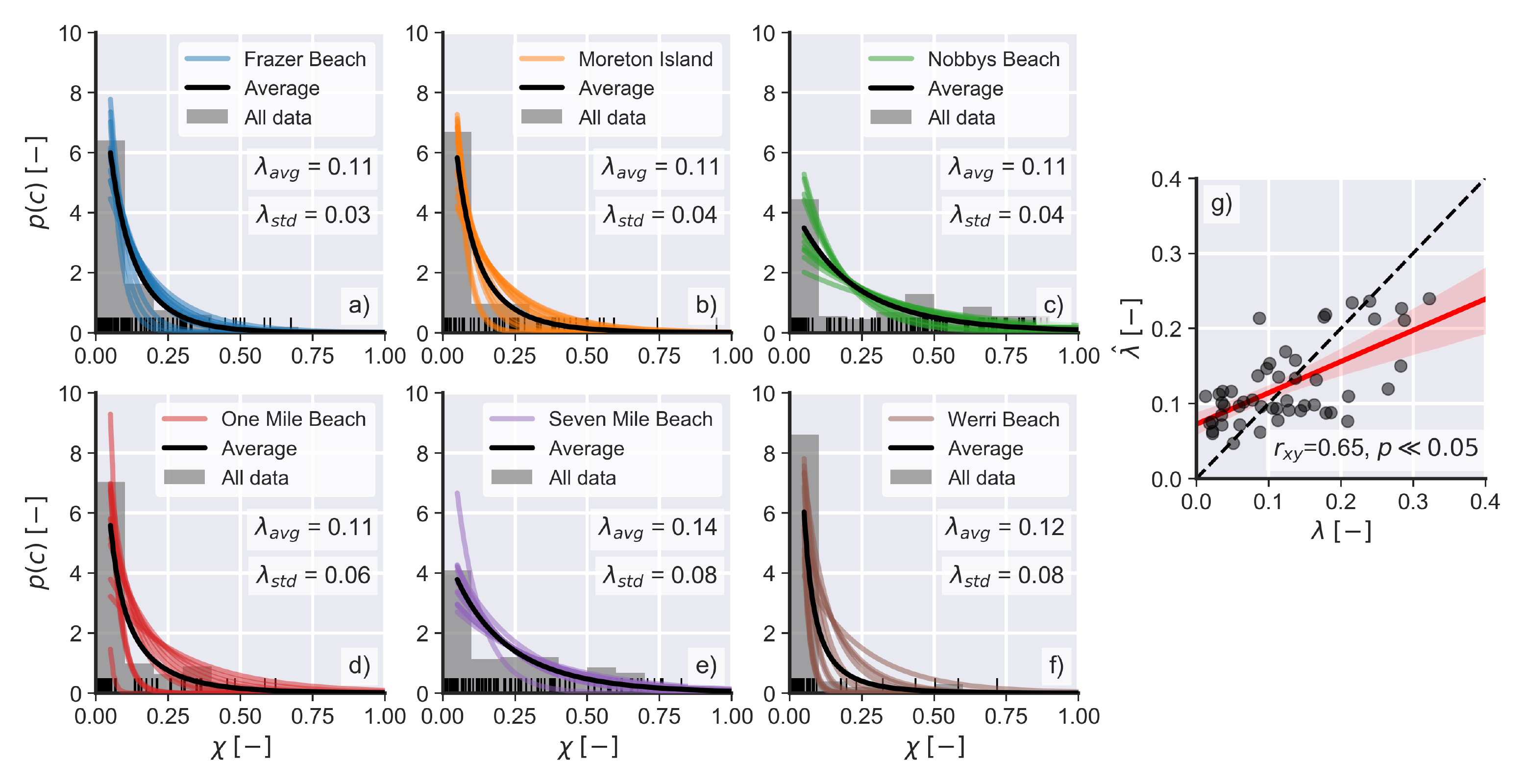}
	\caption{(Caption next page.)}\label{Chap5_Fig_04}
\end{sidewaysfigure}

It is of interest to investigate the most probable location of bore-bore capture because this location may have implications for sediment exchange between the inner surf zone, the swash zone, and the sub-aerial beach \cite{Alsina2018b, Masselink2006}. As such, the nearshore width was defined as the envelope between the surf and swash zone limits (hatched region in Figure \ref{Chap5_Fig_02}) and the location of each bore-bore capture event within this time-varying width was obtained  (Figure \ref{Chap5_Fig_04}). On average, the probability of bore-bore capture at a given nearshore location exponentially decayed such that:

\begin{equation}
p(c) = e^{-\lambda\chi}
\end{equation}

\noindent in which $p(c)$ is the probability density of a bore-bore capture at a given location $\chi$, and $\lambda$ is the exponential decay rate. On average, $\lambda=0.13$, but $\lambda$ varied between beaches and with environmental parameters (see Figure \ref{Chap5_Fig_04}, and Section \ref{Chap_5_Sec_4} for further discussion).

From a coastal engineering point of view, it is also useful to obtain a predictor for the exponential decay rate ($\lambda$) derived from environmental parameters. Several non-dimensional parameterisations to describe $\lambda$ were assessed, but all resulted in lower $r_{xy}$-scores when compared to more sophisticated models. Furthermore, no model based only on offshore parameters was robust enough to predict $\lambda$ accurately ($r_{xy}$-scores $\le0.1$). The best (non-overfitted) parameterisation was the multivariate quadratic relation ($r_{xy}=0.65$):

\begin{equation}
\hat{\lambda} =  0.14-0.05H_{m_{0}}-0.2R+0.69T_{m_{01}}^2-0.02H_{m_{0}}^2+0.07H_{m_{0}}R-0.23R^{2}.\label{Chap_5_Eq_9}
\end{equation}

\noindent in which $R$ is the ratio between IG and SW energy at the time of bore-bore capture averaged over all capture events and was obtained directly from the wavelet transform (see panel d) in Figure \ref{Chap5_Fig_03}). The parameter $R$ is not easily quantified and is rarely available to coastal engineers, therefore, a slightly less robust ($r_{xy}=0.61$) descriptor without the inclusion of this parameter was also obtained:

\begin{equation}
\hat{\lambda} =  0.5-0.04H_{m_{0}}-1.22T_{m_{01}}^2-0.22H_{m_{0}}^2.\label{Chap_5_Eq_10}
\end{equation}

The best non-dimensional single parameter non-quadratic predictor ($r_{xy}=0.45$) was:

\begin{equation}
\hat{\lambda} = 0.23 -0.004\sqrt{H_{m_{0}}L}\label{Chap_5_Eq_11}
\end{equation}

\noindent where $L=\frac{g}{2\pi}T_{m_{01}}^2$. See Section \ref{Chap_5_Sec_4_Sub_1} for further discussion of these predictors.

\addtocounter{figure}{-1}
\begin{figure}
	\caption[Probability density of a bore-bore capture ($p(c)$) occurring at a given normalised cross-shore location ($\chi$) for Frazer Beach, Moreton Island, Nobbys Beach, One Mile Beach, Seven Mile Beach, and Werri Beach.]{Probability density of a bore-bore capture ($p(c)$) occurring at a given normalised cross-shore location ($\chi$), where $\chi=0$ represents the onshore limit of the swash zone and $\chi=1$ represents the offshore limit of the surf zone for each location. Each grey bar represents $p(c)$ for each decile, and the short black lines shows the true distribution of the data. The coloured curves show fits to each 5-minute data run and the black curves show the averaged exponential fit: a) Frazer Beach, b) Moreton Island, c) Nobbys Beach, d) One Mile Beach, e) Seven Mile Beach, and f) Werri Beach. g) Exponential decay rate predictor (Equation \ref{Chap_5_Eq_1}). Each point represents one of the individual coloured curves in a) to e), the red line shows the correlation between the observed ($\lambda$) and predicted ($\hat{\lambda}$) decay rates, the red swath shows the 95\% confidence interval, the black dashed line shows the one-to-one correspondence, and $r_{xy}$ is the Pearson correlation coefficient.}
\end{figure}

\subsection{Infragravity Influence on Bore-Bore Capture}\label{Chap_5_Sec_3_Sub_3}

\begin{figure}[htp]
	\centering
	\includegraphics[width=0.99\textwidth]{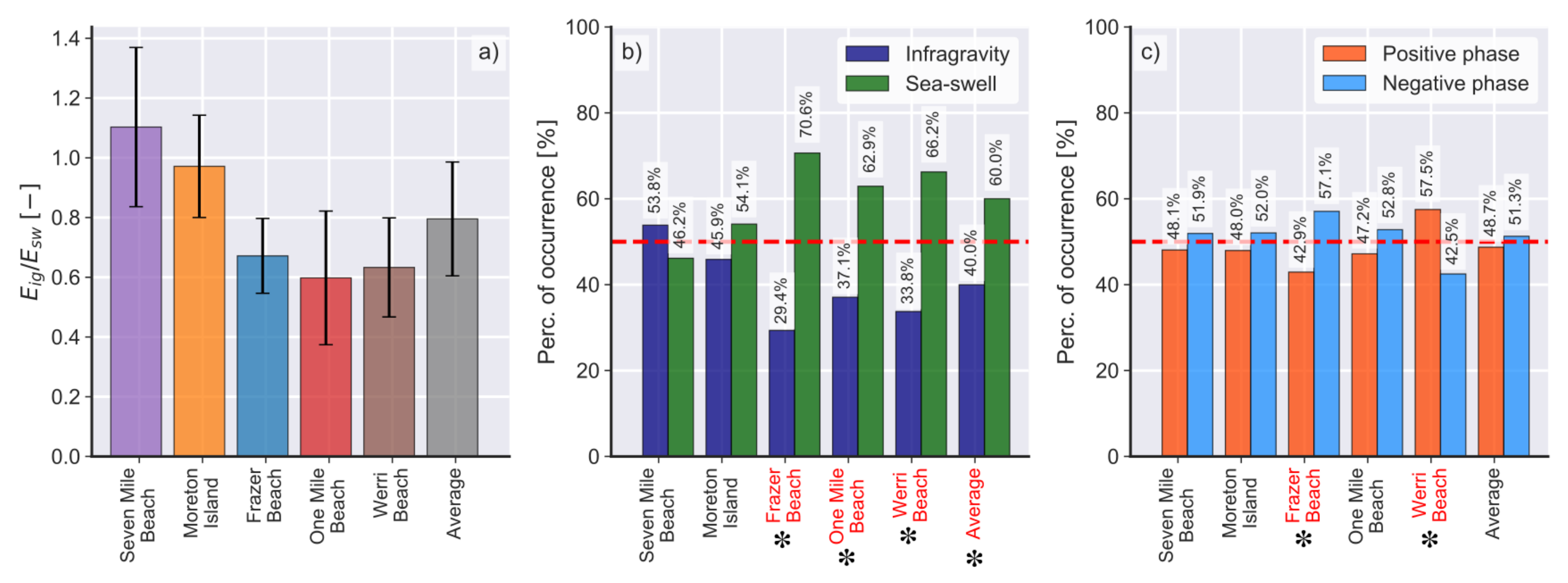}
	\caption[Analysis of infragravity wave influence on bore-bore capture.]{a) Average ratio between IG and SW energy for each beach in which bore-bore capture was observed. b) Percent of occurrence of bore-bore capture events under IG (blue) and SW (green) energy dominance. c) Frequency of occurrence of bore-bore capture events under positive (red) and negative (blue) IG phase. The dashed red lines in b) and c) show the 50\% threshold and the $*$ markers indicates statistically significantly different means using the T-test at the 95\% confidence level. Nobbys Beach data were excluded from this analysis due to the lack of PT data which precluded form calculating $R$.}\label{Chap5_Fig_05}
\end{figure}

The mean capture period results presented in Section \ref{Chap_5_Sec_3_Sub_1} suggested that bore-bore captures do not necessarily occur due to the presence of IG waves (see Table \ref{Chap5_Tab_03}). Thus, it was also of interest to test if bores propagating on the crest of IG waves were more likely to capture other bores than bores propagating on the trough of IG waves. To assess this, bore-bore capture events were correlated to time-dependent SW and IG energy levels (see panel d) in Figure \ref{Chap5_Fig_03}). When the ratio between IG and SW energy was analysed (panel a) in Figure \ref{Chap5_Fig_05}), it was found that, for the majority of the locations, most of the bore-bore capture events occurred under SW dominance ($E_{sw}>E_{ig}$). On average, waves were more likely to merge under SW dominance (60\%) than under IG dominance (40\%) (panel b) in Figure \ref{Chap5_Fig_05}). The only exceptions were SMB and MI where increased IG energy correlated with bore-bore capture events. These results were,  however, not statistically supported by the T-test for two different means ($p>0.05$). Further, the  correlation between bore-bore capture and IG wave phase (panel c) in Figure \ref{Chap5_Fig_05}) showed that bore-bore capture was, on average, equally likely to occur on the trough (51.3\%) or on the crest (48.7\%) of IG waves ($p>0.05$ for the T-test for two different means). The exceptions were WB and FB, in which bore-bore capture events were significantly more likely during the positive (WB) (negative, FB) phase of an IG wave. See Section \ref{Chap_5_Sec_4} for further discussion.

\subsection{Extreme Horizontal Shoreline Excursions}\label{Chap_5_Sec_3_Sub_4}

\begin{figure}[htp]
	\centering
	\includegraphics[width=0.95\textwidth]{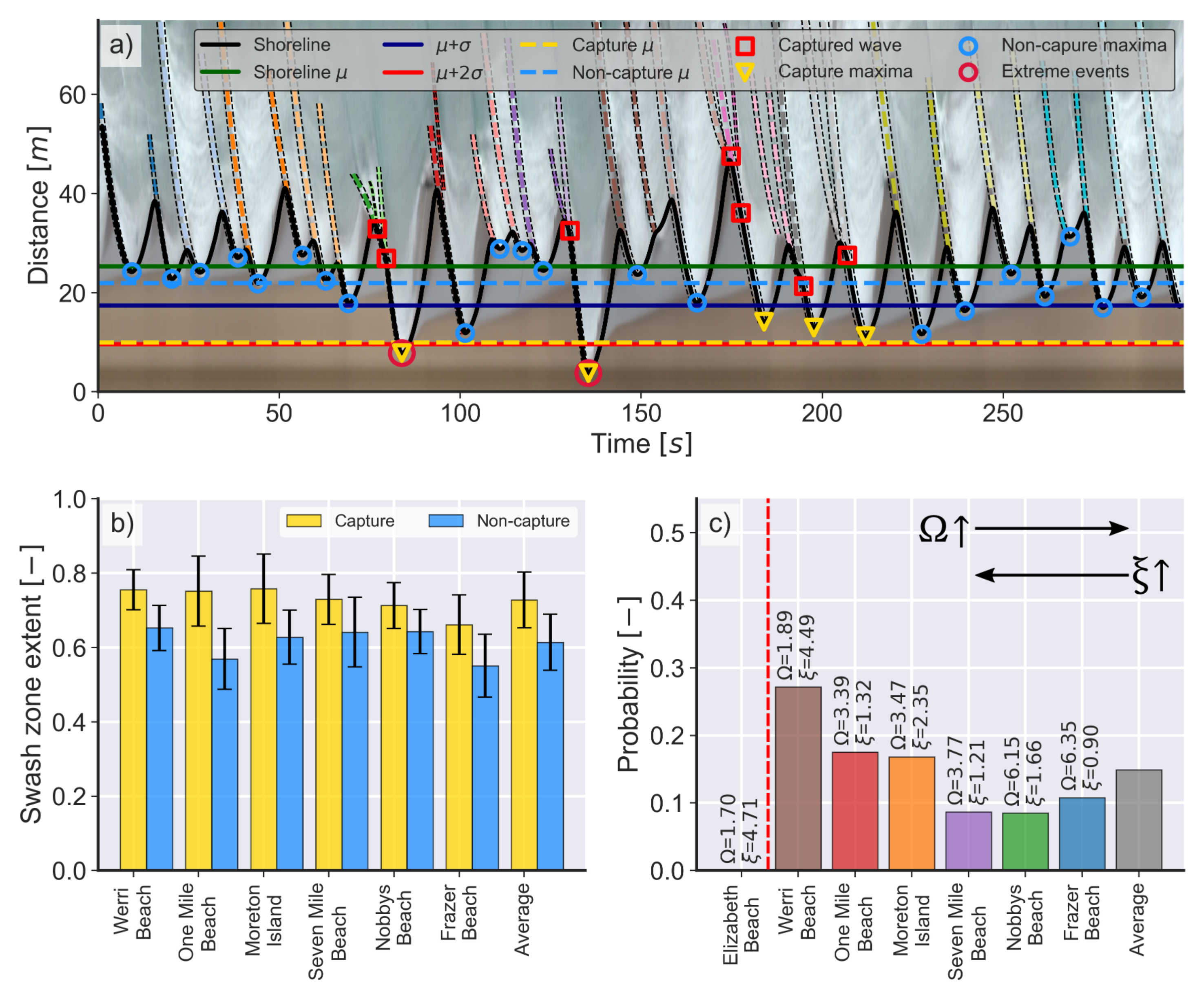}
	\caption[Influence of bore-bore capture on extreme horizontal shoreline maxima.]{Influence of bore-bore capture on extreme horizontal shoreline maxima. a) Example analysis for a subset of WB data. The coloured dashed lines show the tracked waves and their confidence intervals. The thick black line shows the time-varying shoreline position, the green horizontal line shows the mean ($\mu$) shoreline position and the horizontal blue and red lines shows $\mu+\sigma$ and $\mu+2\sigma$, respectively. The red square markers show bore-bore capture events. The red circles show extreme shoreline maxima, the yellow triangles show shoreline maxima derived from bore-bore captures, and the blue circles show shoreline maxima derived from incident waves. The dashed yellow and blue lines show the average shoreline maxima position (over 5-min) driven by bore-bore capture events and incident waves, respectively. b) Averaged (over 5-min) shoreline maxima location for bore-bore capture (yellow) and incident waves (blue). In this panel, swash extent = 0 indicates the surf-swash boundary location and swash extent = 1 indicates the landward limit of the swash zone. c) Probability of a bore-bore capture event generating an extreme shoreline excursion maximum. The red vertical line indicates the observed limit for $\Omega_\infty$ that allowed for bore-bore capture to occur.}\label{Chap5_Fig_06}
\end{figure}

The hypothesis that extreme horizontal shoreline excursions (and consequently extreme runup heights) are directly correlated with bore-bore captures \cite{Garcia-Medina2017} was also tested. Here, an extreme shoreline maxima event was defined as any event that exceeded two standard deviations ($2\sigma$) from the mean shoreline position ($\mu$) (red circles in panel a) in Figure \ref{Chap5_Fig_06}). To quantify if bore-bore capture events led to extreme shoreline maxima, the identified extreme events had to be driven by waves that had undergone bore-bore capture(s) (e.g., the event at 80s in panel a) in Figure 6). In addition, all horizontal shoreline maxima were classed as being derived from a bore-bore capture (yellow triangles in panel a) in Figure \ref{Chap5_Fig_06}) or from an incident wave (blue circles in panel a) in Figure \ref{Chap5_Fig_06}) and the average position for each class were obtained (yellow and blue dashed horizontal lines in panel a) in Figure \ref{Chap5_Fig_06}).

These data showed that $>97\%$ of extreme shoreline maxima were driven by waves that had undergone bore-bore capture. The occurrence of bore-bore capture increased the averaged shoreline maxima position for a given 5-minute intervals by $\approx$10\% across all beaches with no clear trends between the locations (panel b) in Figure \ref{Chap5_Fig_06}). No correlations between the parameters $\Omega_\infty$ \cite{Gourlay1968, Dean1973} or $\xi_\infty$ \cite{Iribarren1949} and the increase in shoreline maxima position caused by the occurrence of bore-bore capture were found. Despite the observation that most extreme shoreline maxima were driven by bore-bore capture, not all bore-bore capture events generated extreme shoreline maxima. In fact, there was a relatively small probability ($<20\%$, on average) of  bore-bore capture driving an extreme shoreline event (panel c) in Figure \ref{Chap5_Fig_06}) as calculated by the number of captures that led to an extreme shoreline maxima divided by the total number of bore-bore captures. 

In contrast, clear relationships between the probability of bore-bore capture driving extreme shoreline maxima and the $\Omega_\infty$ and $\xi_\infty$ parameters were observed (panel c) in Figure \ref{Chap5_Fig_06}): the steeper the beach profile, or the more reflective the beach, the more likely bore-bore capture was to generate an extreme event. Further, the fact that no bore-bore capture was observed at EB suggests that there was an upper (lower) limit for $\xi_\infty$ ($\Omega_\infty$) that allowed for bore-bore capture to occur ($\Omega_\infty \le 1.9$ and $\xi_\infty \ge 4.5$, respectively). The possible causes for this correlation are discussed below in Section \ref{Chap_5_Sec_4}. The IG wave phase was also correlated to the probability of bore-bore capture driving extreme shoreline events but no clear overall trend was observed. For some beaches (FB, MI, and WB) extreme events were more likely during negative IG phases whereas, for others (OMB and SBM), such events were more likely during positive IG phase (not shown).

\section{Discussion}\label{Chap_5_Sec_4}

This paper has presented a novel quantification of bore-bore capture events on natural beaches using machine learning and computer vision techniques applied to data from seven Australian sandy beaches. The results showed that, for beaches where bore-bore capture was observed, there was, on average, a significant probability ($\approxeq$40\%) of any given broken wave being captured by another broken wave in the surf or swash zones.  

The lowest capture probability (25\%) and highest mean capture period (35.6s) were observed at WB most likely due to the higher incident SW wave height and period and the narrower (absolute) surf zone width which, when combined, reduced the probability of bore-bore captures. In contrast, the highest capture probability (42\%) and lowest capture period (17.73s) were observed at FB. Two different factors could explain this result: 1) the low incident wave period in the SW frequency band which may enhance the probability of bore-bore captures via the frequency dispersion mechanism, and 2) the wider (absolute) surf zone that naturally increases the probability of bore-bore captures because the waves have more time and space to interact. Differently to all other analysed beaches, no bore-bore captures were observed at Elizabeth Beach. This likely occurred because of the combination of a high incident wave period (13s), a narrow (absolute) surf zone width, and a very steep beach profile, all of which are likely to inhibit bore-bore capture. Given that the average capture period (24.1s) is very close to the IG cut-off frequency (25s), the results suggest that both IG waves and amplitude and frequency dispersion are equally important for driving bore-bore capture, which is in agreement with the analysis shown in Section \ref{Chap_5_Sec_3_Sub_3}, and with \citeA{Bradshaw1982, Garcia-Medina2017}.

\begin{figure}[htp]
	\centering
	\includegraphics[width=0.95\textwidth]{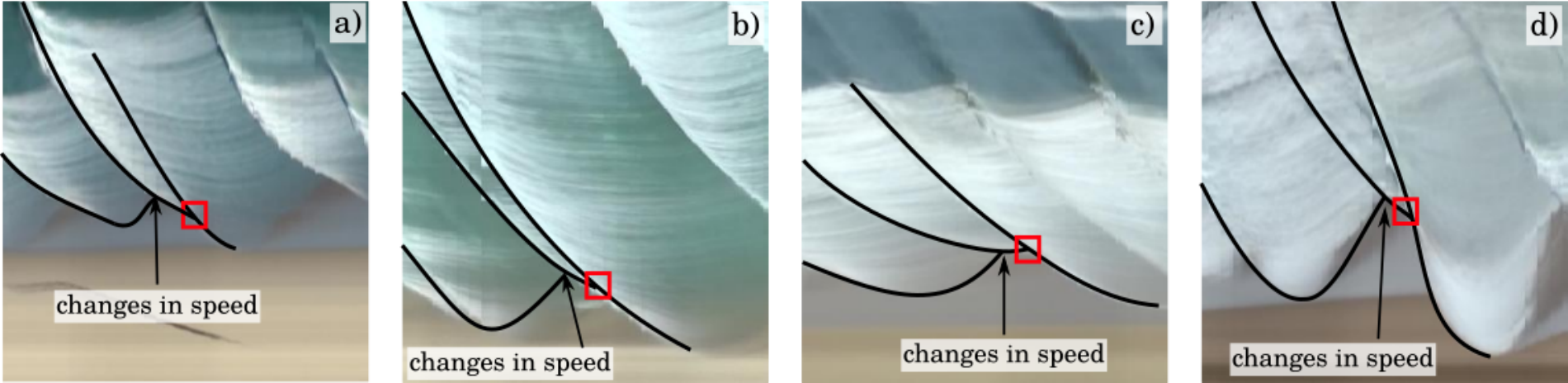}
	\caption[Examples of downrush-uprush interaction causing reductions in the incoming wave speed which and leading to bore-bore capture events.]{Examples of downrush-uprush interaction causing reductions in the incoming wave speed which and leading to bore-bore capture events. a) One Mile Beach, b) Moreton Island, c) Nobbys Beach, and d) Werri Beach. In this figure, all the wave paths and bore-bore capture locations were digitised manually.}\label{Chap5_Fig_07}
\end{figure}

The majority of bore-bore capture events occurred in the landward-most 10\% of the nearshore region. This location was typically observed to occur after the point of bore-collapse of the captured bore, i.e., inside the swash zone, which is equivalent to the swash-wave interaction hydrokinematic region defined by \citeA{Hughes2007}. One mechanism that could explain this location being the observed most probable capturing location are near-bed phenomena (e.g., infiltration and bottom friction), which become significantly more important for energy dissipation in this hydrokinematic region than in other areas of the nearshore \cite{Puleo2001}. These phenomena cause the incoming bore height, and consequently its speed, to decrease rapidly, leading to more frequent bore-bore captures. The interaction between the downrush and the next incoming uprush was also observed to be qualitatively correlated to bore-bore captures on multiple occasions for at least four of the analysed beaches (see Figure \ref{Chap5_Fig_07} for examples). Unfortunately, the precise quantification of changes in the incoming wave speed that were caused due to bottom friction or due to wave-swash interactions and how those changes contribute to driving bore-bore capture was computationally challenging and could not be accomplished using an automated algorithm.

\subsection{Bore-bore Capture Decay Rates}\label{Chap_5_Sec_4_Sub_1}

A predictor for the exponential decay rate ($\lambda$) of the probability of a bore-bore capture occurring at a given normalised nearshore location ($\chi$) was presented in Section \ref{Chap_5_Sec_2}. The analysis of the coefficients of the full predictive model (Equation \ref{Chap_5_Eq_9}) shows that the most relevant term in the regression is the square of the wave period, and that the square of ratio between IG and SW energy is relatively important for the model. However, when this ratio ($R=E_{ig}/E_{sw}$) is not included in the model, there is only a 4\% reduction in $r_{xy}$  which shows that, for predictive purposes, there is no major dependence of $\lambda$ on $R$. For the model without $R$ (Equation \ref{Chap_5_Eq_10}), a strong dependence of $\lambda$ on the square of the wave period was again observed, indicating that frequency dispersion, which is a function of $T$ via the linear wave dispersion equation, is key for $\lambda$ predictions.

The best linear model (Equation \ref{Chap_5_Eq_11}) is similar to the runup descriptors seen in \citeA{Nielsen1991, Stockdon2006} with a dependence on $\sqrt{H_{m0}L}$. This predictor is also a direct function of the wave period, which further correlates $\lambda$ to $T$. The inclusion of a beach slope parameter (e.g., surf zone beach slope, swash zone beach slope, or normalised beach slope) in any of the parametrisations did not improve the predictions. In fact, the inclusion of an extra parameter only increased the models' complexity without any statistically significant improvement when metrics such as the Akaike Information Criterion \cite{Akaike1974} were used to assess the predictors. 

\subsection{IG Energy Influence on Bore-Bore Capture}\label{Chap_5_Sec_4_Sub_2}

The data analysed has shown that bore-bore capture was, on average, equally likely to occur on the crest (positive phase) and on the trough (negative phase) of IG waves, and that a surf zone dominated by IG energy was no more likely to have bore-bore capture events than a surf zone dominated by SW energy. These results are complementary to the findings of previous studies that exclusively focused on the effect of IG waves on bore-bore capture (e.g., \citeA{Tissier2015, VanDongeren2007}). These previous studies were based on numerical modelling of laboratory experiments on planar sloping beaches ($\beta=\{0.02, 0.01\}$) under bi-chromatic wave forcing, which do not typically represent the random wave field observed on natural beaches. Furthermore, as identified by \citeA{Baldock2012}, such gentle beach slopes may not be representative of the IG dynamics observed on natural beaches. The present analyses suggest that, under the broad wave spectra present in natural surf zones, the differences in local depth due to IG wave modulation were not sufficiently large to be the likely driver of bore-bore capture. This further suggests that wave dispersion is the likely underlying mechanism, such that if the differences in bore height or period between two consecutive bores are sufficiently large to allow for the faster moving bore to capture the slower one, the capture will occur regardless of the IG wave phase.

The averaged null influence of IG phase on bore-bore capture (see panel c) in Figure \ref{Chap5_Fig_05}) is likely due to the fact the IG wave amplitude is often at least one order of magnitude less than the SW amplitude \cite{Battjes2004, Baldock2012, Bertin2018b}, which can be the same order of magnitude as the differences in wave heights within the wave group structure. This result is also consistent with the observation by \citeA{Baldock2012} that the IG released after short-wave breaking often quickly decays and, therefore, is not typically large enough to overcome individual wave height differences within the wave group. However, a more detailed field dataset is required to accurately quantify the effects of the wave group structure and wave dispersion on bore-bore capture. This should be focus of future research.

One limitation of the analysis developed here was that only the bulk contribution of IG waves was considered. A shoreline-reflected outgoing IG wave may also be present on natural beaches and have influence on short wave dynamics \cite{Battjes2004, Contardo2013}. The amplitude of this wave is often much smaller than the incoming IG wave (e.g., \citeA{Battjes2004}, their Figure 4) and so is unlikely to significantly change the patterns observed here, especially given that the incoming IG wave seemed not to significantly affect bore-bore capture. Another limitation is that it was impossible to precisely quantify the exact threshold at which frequency and amplitude dispersion overcame IG modulation. Due to the random characteristics of the wave field, bore-bore capture can and does occur throughout the full extension of the surf and swash zones, but the PT locations were discrete and rarely coincided with the identified capture events. This limitation could be overcome by using LIDAR based methods which record continuous cross-shore water surface elevations.

\subsection{Effect of Storm Conditions}\label{Chap_5_Sec_4_Sub_3}

The effect of storm conditions on bore-bore capture was also tested. It is frequently assumed that IG wave energy is dominant under storm conditions \cite{Russell1993, Bakker2016a} and, therefore, could result in enhanced bore-bore capture via the mechanism described by \citeA{Tissier2015}. The probability of bore-bore capture was higher in the storm conditions observed at NB when compared to the other locations (Table \ref{Chap5_Tab_03}), which could be consistent with the observation by \citeA{Russell1993}, or could be due to the broader spectral width typically associated with storms which would lead to enhanced wave dispersion \cite{holthuijsen2010}. At NB, the most probable location for bore-bore capture remained in the landward-most 10\% of the nearshore region, which could be due to near-bed or reflection phenomena as discussed above (e.g., panel c) in Figure \ref{Chap5_Fig_07}). There was, however, a significant increase in $p(c)$ in the outer surf zone (panel c) in Figure \ref{Chap5_Fig_04}). This is in contrast to the mechanism described by \cite{Tissier2015} as the amplitude of the IG wave is at a minimum in this region \cite{Bakker2014,VanDongeren2007}. Furthermore, \citeA{Garcia-Medina2017} showed that, in the outer surf zone, amplitude dispersion is more likely to lead to bore-bore capture than IG modulation. Therefore, it is more likely that the bore-bore capture dynamics observed at NB were due to amplitude and frequency dispersion than to IG wave modulation. Further, under the storm conditions observed at NB, the decay rate parameter ($\lambda$) also increased significantly, which indicates that, under these conditions, the predictors presented in Section \ref{Chap_5_Sec_3_Sub_2} must be used with caution.

Finally, from the results in Figure \ref{Chap5_Fig_06}, the presence of storm conditions did not seem to significantly increase  average shoreline maxima  positions driven by bore-bore capture when compared to the incident wave average in the other locations (panel b) in Figure \ref{Chap5_Fig_06}), nor there was an increase in the probability of bore-bore capture driving shoreline extrema (panel c) in Figure \ref{Chap5_Fig_06}). These results suggest that the increase in shoreline maxima driven by bore-bore capture is invariant under storm conditions.


\subsection{Implications for Future Studies on Runup Dynamics}\label{Chap_5_Sec_4_Sub_4}

The results of the correlation between bore-bore capture and extreme shoreline maxima showed that only a small proportion of bore-bore capture events resulted in extreme shoreline maxima, however, nearly all ($>97\%$) extreme shoreline maxima were directly driven by bore-bore capture. The probability of a bore-bore capture event generating an extreme shoreline maxima was closely correlated to $\Omega_\infty$ and $\xi_\infty$: the more reflective (or steeper) the beach, the more likely a bore-bore capture event was to generate an extreme event up to a minimum (maximum) value of $\Omega_\infty$ ($\xi_\infty$) beyond which, no bore-bore capture was observed. For the data analysed here, the cut-off beyond which no bore-bore capture was observed was $\xi_\infty \ge 4.5$ and $\Omega_\infty \le 1.9$ at EB, i.e., for near-reflective conditions ($\Omega_\infty \le 2$) and plunging/surging breakers ($\xi_\infty >3.3$) \cite{Galvin1968, Wright1984}.

From the present data, it is possible to infer that bore-bore capture punctuates the shoreward moment flux (the net incoming momentum flux is constant), leading to higher-than-normal shoreline excursions due to mass and momentum conservation; particularly if the capture occurs after the point of bore collapse of the leading wave. This is further supported from the fact that bore-bore capture is more likely to generate extreme shoreline maxima for steeper beach profiles, which has shown to be the case via numerical modelling for bichromatic waves \cite{Garcia-Medina2017}.

The occurrence of bore-bore capture increased the averaged shoreline maxima position by $\approx$10\% when compared to the averaged shoreline maxima position for incident waves only (panel b) in Figure \ref{Chap5_Fig_06}). This result may explain some of the variability seen in parametric runup models that exclude this variable (e.g., see \citeA{Atkinson2017} M2 model). Further, the results presented here do not seem to support the infragravity-based swash model often assumed to be the main driver for extreme runup on natural beaches \cite{Guza1982, Holman1986, Nielsen1991, Holland1995, Stockdon2006, Plant2015}. This suggests that more focus should be given to understanding the wave-group structure, wave-wave, and wave-swash interactions to better understand extreme swash zone dynamics. Given that the standard approach for parametric runup modelling cannot account for the effects of bore-bore capture directly, i.e., these models are not driven by physical processes, the increase in the shoreline maxima due to bore-bore capture could be parameterised via the Iribarren number and included in the models directly. It has previously been shown that, despite parametric runup models not being driven by physical process, the inclusion of parameterised physical processes (e.g., wave setup) can increase the models' accuracy \cite{Stockdon2006, Holman1986}. This is, however, beyond the scope of this work and should be focus of future research.

\section{Conclusion}\label{Chap_5_Sec_5}

Wave tracking applied to data collected at seven natural beaches allowed for novel  quantification of bore-bore capture, which is shown to be a common feature of natural nearshore environments. For beaches where bore-bore capture was observed, there was a significant probability ($\approxeq$40\%) of a bore being captured by another bore in the surf or swash zones. In addition, the present data show a absence of bore-bore capture for conditions with $\xi_\infty \ge 4.5$ and $\Omega_\infty \le 1.9$, i.e., under near-reflective conditions. In contrast to previous laboratory and modelling studies, bore-bore capture was not found to be predominantly controlled by the presence of infragravity waves. Consequently, amplitude and frequency dispersion are inferred to be the main physical drivers. The majority of bore-bore capture events occurred after the point of bore collapse of the captured bore, i.e., well inside the swash zone in the swash-wave interaction hydrokinematic region. This most frequent capture location suggests that near-bed phenomena, surf-swash and swash-swash interactions, and swash-based reflection become significantly important for the dispersion relation in this hydrokinematic region. Finally, a relatively small percentage of bore-bore capture events led to extreme shoreline maxima ($<$20\% on average), however, virtually all ($>$97\%) extreme shoreline maxima were directly driven by bore-bore capture events. Ultimately, the results presented here show that bore-bore capture is an important phenomena driving extreme shoreline dynamics and, therefore, should be directly accounted for in future considerations of extreme runup modelling.

\acknowledgments

The field work (2014 experiments) were funded by a University of Newcastle Faculty of Science and I.T. Strategic Initiatives Research Fund Grant 2014 to HEP. The authors are grateful to Alex Atkinson, Andrew Magee, Annette Burke, Emily Kirk, Daniel Harris, David Hanslow, Kaya Wilson, Madeleine Broadfoot, Michael Hughes, Mike Kinsela, Murray Kendall, Rachael Grant, Rebecca Hamilton, Samantha Clarke, and Tom Donaldson-Brown who assisted with the field data collection. CES is funded by a University of Newcastle Research Degree Scholarship (UNRS) 5050UNRS and a Central \& Faculty scholarship. The pressure transducer and video cameras used in the 2014 experiments were kindly lent by Tom Baldock from the University of Queensland. The sediment data were kindly provided by Professor Andrew Short from the University of Sydney. The authors are also thankful to the Academic Research Computing Support Team, particularly Aaron Scott, at the University of Newcastle for support with the I.T. infrastructure on which all video data pre-processing and machine-learning development were undertaken, and to Bas Hoonhout who helped providing the original image rectification codes.

\section*{Data Availability}

Code and data for this paper are available at \url{https://github.com/caiostringari/BBC-JGR-Oceans}. Note that due to repository size restrictions, raw data imagery could not be uploaded. Video imagery data is, however, fully available upon request.

%

%
\bibliography{library.bib}

\begin{thebibliography}{}

\bibitem [\protect \citeauthoryear {%
Aagaard%
\ \BBA {} Holm%
}{%
Aagaard%
\ \BBA {} Holm%
}{%
{\protect \APACyear {1989}}%
}]{%
Aagaard1989}
\APACinsertmetastar {%
Aagaard1989}%
\begin{APACrefauthors}%
Aagaard, T.%
\BCBT {}\ \BBA {} Holm, J.%
\end{APACrefauthors}%
\unskip\
\newblock
\APACrefYearMonthDay{1989}{}{}.
\newblock
{\BBOQ}\APACrefatitle {{Digitization of Wave Run-up Using Video Records}}
  {{Digitization of Wave Run-up Using Video Records}}.{\BBCQ}
\newblock
\APACjournalVolNumPages{Journal of Coastal Research}{5}{3}{547--551}.
\PrintBackRefs{\CurrentBib}

\bibitem [\protect \citeauthoryear {%
Addison%
, Watson%
\BCBL {}\ \BBA {} Feng%
}{%
Addison%
\ \protect \BOthers {.}}{%
{\protect \APACyear {2002}}%
}]{%
Addison2002}
\APACinsertmetastar {%
Addison2002}%
\begin{APACrefauthors}%
Addison, P\BPBI S.%
, Watson, J\BPBI N.%
\BCBL {}\ \BBA {} Feng, T.%
\end{APACrefauthors}%
\unskip\
\newblock
\APACrefYearMonthDay{2002}{}{}.
\newblock
{\BBOQ}\APACrefatitle {{Low-oscillation complex wavelets}} {{Low-oscillation
  complex wavelets}}.{\BBCQ}
\newblock
\APACjournalVolNumPages{Journal of Sound and Vibration}{254}{4}{733--762}.
\newblock
\begin{APACrefDOI} \doi{10.1006/jsvi.2001.4119} \end{APACrefDOI}
\PrintBackRefs{\CurrentBib}

\bibitem [\protect \citeauthoryear {%
Akaike%
}{%
Akaike%
}{%
{\protect \APACyear {1974}}%
}]{%
Akaike1974}
\APACinsertmetastar {%
Akaike1974}%
\begin{APACrefauthors}%
Akaike, H.%
\end{APACrefauthors}%
\unskip\
\newblock
\APACrefYearMonthDay{1974}{}{}.
\newblock
{\BBOQ}\APACrefatitle {{A New Look at the Statistical Model Identification}}
  {{A New Look at the Statistical Model Identification}}.{\BBCQ}
\newblock
\APACjournalVolNumPages{IEEE Transactions on Automatic
  Control}{19}{6}{716--723}.
\newblock
\begin{APACrefDOI} \doi{10.1109/TAC.1974.1100705} \end{APACrefDOI}
\PrintBackRefs{\CurrentBib}

\bibitem [\protect \citeauthoryear {%
Almar%
, Larnier%
, Castelle%
, Scott%
\BCBL {}\ \BBA {} Floc'h%
}{%
Almar%
\ \protect \BOthers {.}}{%
{\protect \APACyear {2016}}%
}]{%
Almar2016}
\APACinsertmetastar {%
Almar2016}%
\begin{APACrefauthors}%
Almar, R.%
, Larnier, S.%
, Castelle, B.%
, Scott, T.%
\BCBL {}\ \BBA {} Floc'h, F.%
\end{APACrefauthors}%
\unskip\
\newblock
\APACrefYearMonthDay{2016}{}{}.
\newblock
{\BBOQ}\APACrefatitle {{On the use of the Radon transform to estimate longshore
  currents from video imagery}} {{On the use of the Radon transform to estimate
  longshore currents from video imagery}}.{\BBCQ}
\newblock
\APACjournalVolNumPages{Coastal Engineering}{114}{}{301--308}.
\newblock
\begin{APACrefURL} \url{http://dx.doi.org/10.1016/j.coastaleng.2016.04.016}
  \end{APACrefURL}
\newblock
\begin{APACrefDOI} \doi{10.1016/j.coastaleng.2016.04.016} \end{APACrefDOI}
\PrintBackRefs{\CurrentBib}

\bibitem [\protect \citeauthoryear {%
Alsina%
, van~der Zanden%
, C{\'{a}}ceres%
\BCBL {}\ \BBA {} Ribberink%
}{%
Alsina%
\ \protect \BOthers {.}}{%
{\protect \APACyear {2018}}%
}]{%
Alsina2018b}
\APACinsertmetastar {%
Alsina2018b}%
\begin{APACrefauthors}%
Alsina, J\BPBI M.%
, van~der Zanden, J.%
, C{\'{a}}ceres, I.%
\BCBL {}\ \BBA {} Ribberink, J\BPBI S.%
\end{APACrefauthors}%
\unskip\
\newblock
\APACrefYearMonthDay{2018}{}{}.
\newblock
{\BBOQ}\APACrefatitle {{The influence of wave groups and wave-swash
  interactions on sediment transport and bed evolution in the swash zone}}
  {{The influence of wave groups and wave-swash interactions on sediment
  transport and bed evolution in the swash zone}}.{\BBCQ}
\newblock
\APACjournalVolNumPages{Coastal Engineering}{140}{May}{23--42}.
\newblock
\begin{APACrefURL} \url{https://doi.org/10.1016/j.coastaleng.2018.06.005}
  \end{APACrefURL}
\newblock
\begin{APACrefDOI} \doi{10.1016/j.coastaleng.2018.06.005} \end{APACrefDOI}
\PrintBackRefs{\CurrentBib}

\bibitem [\protect \citeauthoryear {%
Atkinson%
\ \protect \BOthers {.}}{%
Atkinson%
\ \protect \BOthers {.}}{%
{\protect \APACyear {2017}}%
}]{%
Atkinson2017}
\APACinsertmetastar {%
Atkinson2017}%
\begin{APACrefauthors}%
Atkinson, A\BPBI L.%
, Power, H\BPBI E.%
, Moura, T.%
, Hammond, T.%
, Callaghan, D\BPBI P.%
\BCBL {}\ \BBA {} Baldock, T\BPBI E.%
\end{APACrefauthors}%
\unskip\
\newblock
\APACrefYearMonthDay{2017}{}{}.
\newblock
{\BBOQ}\APACrefatitle {{Assessment of runup predictions by empirical models on
  non-truncated beaches on the south-east Australian coast}} {{Assessment of
  runup predictions by empirical models on non-truncated beaches on the
  south-east Australian coast}}.{\BBCQ}
\newblock
\APACjournalVolNumPages{Coastal Engineering}{119}{March 2016}{15--31}.
\newblock
\begin{APACrefURL} \url{http://dx.doi.org/10.1016/j.coastaleng.2016.10.001}
  \end{APACrefURL}
\newblock
\begin{APACrefDOI} \doi{10.1016/j.coastaleng.2016.10.001} \end{APACrefDOI}
\PrintBackRefs{\CurrentBib}

\bibitem [\protect \citeauthoryear {%
Baldock%
}{%
Baldock%
}{%
{\protect \APACyear {2012}}%
}]{%
Baldock2012}
\APACinsertmetastar {%
Baldock2012}%
\begin{APACrefauthors}%
Baldock, T\BPBI E.%
\end{APACrefauthors}%
\unskip\
\newblock
\APACrefYearMonthDay{2012}{}{}.
\newblock
{\BBOQ}\APACrefatitle {{Dissipation of incident forced long waves in the surf
  zone-Implications for the concept of "bound" wave release at short wave
  breaking}} {{Dissipation of incident forced long waves in the surf
  zone-Implications for the concept of "bound" wave release at short wave
  breaking}}.{\BBCQ}
\newblock
\APACjournalVolNumPages{Coastal Engineering}{60}{1}{276--285}.
\newblock
\begin{APACrefURL} \url{http://dx.doi.org/10.1016/j.coastaleng.2011.11.002}
  \end{APACrefURL}
\newblock
\begin{APACrefDOI} \doi{10.1016/j.coastaleng.2011.11.002} \end{APACrefDOI}
\PrintBackRefs{\CurrentBib}

\bibitem [\protect \citeauthoryear {%
Battjes%
, Bakkenes%
, Janssen%
\BCBL {}\ \BBA {} Dongeren%
}{%
Battjes%
\ \protect \BOthers {.}}{%
{\protect \APACyear {2004}}%
}]{%
Battjes2004}
\APACinsertmetastar {%
Battjes2004}%
\begin{APACrefauthors}%
Battjes, J\BPBI A.%
, Bakkenes, H\BPBI J.%
, Janssen, T\BPBI T.%
\BCBL {}\ \BBA {} Dongeren, A\BPBI R\BPBI V.%
\end{APACrefauthors}%
\unskip\
\newblock
\APACrefYearMonthDay{2004}{}{}.
\newblock
{\BBOQ}\APACrefatitle {{Shoaling of subharmonic gravity waves}} {{Shoaling of
  subharmonic gravity waves}}.{\BBCQ}
\newblock
\APACjournalVolNumPages{Journal Of Geophysical Research}{109}{}{1--15}.
\newblock
\begin{APACrefDOI} \doi{10.1029/2003JC001863} \end{APACrefDOI}
\PrintBackRefs{\CurrentBib}

\bibitem [\protect \citeauthoryear {%
Bertin%
\ \protect \BOthers {.}}{%
Bertin%
\ \protect \BOthers {.}}{%
{\protect \APACyear {2018}}%
}]{%
Bertin2018b}
\APACinsertmetastar {%
Bertin2018b}%
\begin{APACrefauthors}%
Bertin, X.%
, de Bakker, A.%
, van Dongeren, A.%
, Coco, G.%
, Andr{\'{e}}, G.%
, Ardhuin, F.%
\BDBL {}Tissier, M.%
\end{APACrefauthors}%
\unskip\
\newblock
\APACrefYearMonthDay{2018}{}{}.
\newblock
{\BBOQ}\APACrefatitle {{Infragravity waves: From driving mechanisms to
  impacts}} {{Infragravity waves: From driving mechanisms to impacts}}.{\BBCQ}
\newblock
\APACjournalVolNumPages{Earth-Science Reviews}{177}{June 2017}{774--799}.
\newblock
\begin{APACrefURL} \url{https://doi.org/10.1016/j.earscirev.2018.01.002}
  \end{APACrefURL}
\newblock
\begin{APACrefDOI} \doi{10.1016/j.earscirev.2018.01.002} \end{APACrefDOI}
\PrintBackRefs{\CurrentBib}

\bibitem [\protect \citeauthoryear {%
Bowen%
\ \BBA {} Guza%
}{%
Bowen%
\ \BBA {} Guza%
}{%
{\protect \APACyear {1978}}%
}]{%
Bowen1978}
\APACinsertmetastar {%
Bowen1978}%
\begin{APACrefauthors}%
Bowen, J.%
\BCBT {}\ \BBA {} Guza, R.%
\end{APACrefauthors}%
\unskip\
\newblock
\APACrefYearMonthDay{1978}{}{}.
\newblock
{\BBOQ}\APACrefatitle {{Edge Waves and Surf Beat}} {{Edge Waves and Surf
  Beat}}.{\BBCQ}
\newblock
\APACjournalVolNumPages{Journal of Geophysical Research}{83}{7}{}.
\PrintBackRefs{\CurrentBib}

\bibitem [\protect \citeauthoryear {%
Bradshaw%
}{%
Bradshaw%
}{%
{\protect \APACyear {1982}}%
}]{%
Bradshaw1982}
\APACinsertmetastar {%
Bradshaw1982}%
\begin{APACrefauthors}%
Bradshaw, M.%
\end{APACrefauthors}%
\unskip\
\newblock
\APACrefYearMonthDay{1982}{}{}.
\newblock
\APACrefbtitle {{Bores and Swash on Natural Beaches}} {{Bores and Swash on
  Natural Beaches}}\ \APACbVolEdTR{}{\BTR{}}.
\newblock
\APACaddressInstitution{Syd}{University of Sydney}.
\PrintBackRefs{\CurrentBib}

\bibitem [\protect \citeauthoryear {%
Brocchini%
\ \BBA {} Baldock%
}{%
Brocchini%
\ \BBA {} Baldock%
}{%
{\protect \APACyear {2008}}%
}]{%
Brocchini2008}
\APACinsertmetastar {%
Brocchini2008}%
\begin{APACrefauthors}%
Brocchini, M.%
\BCBT {}\ \BBA {} Baldock, T\BPBI E.%
\end{APACrefauthors}%
\unskip\
\newblock
\APACrefYearMonthDay{2008}{}{}.
\newblock
{\BBOQ}\APACrefatitle {{Recent advances in modeling swash zone dynamics:
  Influence of surf-swash interaction on nearshore hydrodynamics and
  morphodynamics}} {{Recent advances in modeling swash zone dynamics: Influence
  of surf-swash interaction on nearshore hydrodynamics and
  morphodynamics}}.{\BBCQ}
\newblock
\APACjournalVolNumPages{Reviews of Geophysics}{46}{3}{1--21}.
\newblock
\begin{APACrefDOI} \doi{10.1029/2006RG000215} \end{APACrefDOI}
\PrintBackRefs{\CurrentBib}

\bibitem [\protect \citeauthoryear {%
Chard{\'{o}}n-Maldonado%
, Pintado-Pati{\~{n}}o%
\BCBL {}\ \BBA {} Puleo%
}{%
Chard{\'{o}}n-Maldonado%
\ \protect \BOthers {.}}{%
{\protect \APACyear {2016}}%
}]{%
Chardon-Maldonado2016}
\APACinsertmetastar {%
Chardon-Maldonado2016}%
\begin{APACrefauthors}%
Chard{\'{o}}n-Maldonado, P.%
, Pintado-Pati{\~{n}}o, J\BPBI C.%
\BCBL {}\ \BBA {} Puleo, J\BPBI A.%
\end{APACrefauthors}%
\unskip\
\newblock
\APACrefYearMonthDay{2016}{}{}.
\newblock
{\BBOQ}\APACrefatitle {{Advances in swash-zone research: Small-scale
  hydrodynamic and sediment transport processes}} {{Advances in swash-zone
  research: Small-scale hydrodynamic and sediment transport processes}}.{\BBCQ}
\newblock
\APACjournalVolNumPages{Coastal Engineering}{115}{}{8--25}.
\newblock
\begin{APACrefURL} \url{http://dx.doi.org/10.1016/j.coastaleng.2015.10.008}
  \end{APACrefURL}
\newblock
\begin{APACrefDOI} \doi{10.1016/j.coastaleng.2015.10.008} \end{APACrefDOI}
\PrintBackRefs{\CurrentBib}

\bibitem [\protect \citeauthoryear {%
Contardo%
\ \BBA {} Symonds%
}{%
Contardo%
\ \BBA {} Symonds%
}{%
{\protect \APACyear {2013}}%
}]{%
Contardo2013}
\APACinsertmetastar {%
Contardo2013}%
\begin{APACrefauthors}%
Contardo, S.%
\BCBT {}\ \BBA {} Symonds, G.%
\end{APACrefauthors}%
\unskip\
\newblock
\APACrefYearMonthDay{2013}{}{}.
\newblock
{\BBOQ}\APACrefatitle {{Infragravity response to variable wave forcing in the
  nearshore}} {{Infragravity response to variable wave forcing in the
  nearshore}}.{\BBCQ}
\newblock
\APACjournalVolNumPages{Journal of Geophysical Research:
  Oceans}{118}{12}{7095--7106}.
\newblock
\begin{APACrefDOI} \doi{10.1002/2013JC009430} \end{APACrefDOI}
\PrintBackRefs{\CurrentBib}

\bibitem [\protect \citeauthoryear {%
Dean%
}{%
Dean%
}{%
{\protect \APACyear {1973}}%
}]{%
Dean1973}
\APACinsertmetastar {%
Dean1973}%
\begin{APACrefauthors}%
Dean, R\BPBI G.%
\end{APACrefauthors}%
\unskip\
\newblock
\APACrefYearMonthDay{1973}{}{}.
\newblock
{\BBOQ}\APACrefatitle {{Heuristic Models of Sand Transport in the Surf Zone}}
  {{Heuristic Models of Sand Transport in the Surf Zone}}.{\BBCQ}
\newblock
\BIn{} \APACrefbtitle {First Australian Conference on Coastal Engineering}
  {First australian conference on coastal engineering}\ (\BPGS\ 215--221).
\newblock
\APACaddressPublisher{Sydney}{}.
\PrintBackRefs{\CurrentBib}

\bibitem [\protect \citeauthoryear {%
de Bakker%
, Brinkkemper%
, Steen%
, Tissier%
\BCBL {}\ \BBA {} Ruessink%
}{%
de Bakker%
\ \protect \BOthers {.}}{%
{\protect \APACyear {2016}}%
}]{%
Bakker2016a}
\APACinsertmetastar {%
Bakker2016a}%
\begin{APACrefauthors}%
de Bakker, A\BPBI T\BPBI M.%
, Brinkkemper, J\BPBI A.%
, Steen, F\BPBI V\BPBI D.%
, Tissier, M\BPBI F\BPBI S.%
\BCBL {}\ \BBA {} Ruessink, B\BPBI G.%
\end{APACrefauthors}%
\unskip\
\newblock
\APACrefYearMonthDay{2016}{}{}.
\newblock
{\BBOQ}\APACrefatitle {{Cross-shore sand transport by infragravity waves as a
  function of beach steepness}} {{Cross-shore sand transport by infragravity
  waves as a function of beach steepness}}.{\BBCQ}
\newblock
\APACjournalVolNumPages{Journal Of Geophysical Research: Earth
  Surface}{121}{}{1786--1799}.
\newblock
\begin{APACrefDOI} \doi{10.1002/2016JF003878.Abstract} \end{APACrefDOI}
\PrintBackRefs{\CurrentBib}

\bibitem [\protect \citeauthoryear {%
de Bakker%
, Tissier%
\BCBL {}\ \BBA {} Ruessink%
}{%
de Bakker%
\ \protect \BOthers {.}}{%
{\protect \APACyear {2014}}%
}]{%
Bakker2014}
\APACinsertmetastar {%
Bakker2014}%
\begin{APACrefauthors}%
de Bakker, A\BPBI T\BPBI M.%
, Tissier, M\BPBI F\BPBI S.%
\BCBL {}\ \BBA {} Ruessink, B\BPBI G.%
\end{APACrefauthors}%
\unskip\
\newblock
\APACrefYearMonthDay{2014}{}{}.
\newblock
{\BBOQ}\APACrefatitle {{Shoreline dissipation of infragravity waves}}
  {{Shoreline dissipation of infragravity waves}}.{\BBCQ}
\newblock
\APACjournalVolNumPages{Continental Shelf Research}{72}{}{73--82}.
\newblock
\begin{APACrefURL} \url{http://dx.doi.org/10.1016/j.csr.2013.11.013}
  \end{APACrefURL}
\newblock
\begin{APACrefDOI} \doi{10.1016/j.csr.2013.11.013} \end{APACrefDOI}
\PrintBackRefs{\CurrentBib}

\bibitem [\protect \citeauthoryear {%
de Moura%
\ \BBA {} Baldock%
}{%
de Moura%
\ \BBA {} Baldock%
}{%
{\protect \APACyear {2017}}%
}]{%
DeMoura2017}
\APACinsertmetastar {%
DeMoura2017}%
\begin{APACrefauthors}%
de Moura, T\BPBI G\BPBI R.%
\BCBT {}\ \BBA {} Baldock, T\BPBI E.%
\end{APACrefauthors}%
\unskip\
\newblock
\APACrefYearMonthDay{2017}{}{}.
\newblock
{\BBOQ}\APACrefatitle {{Remote sensing of the correlation between breakpoint
  oscillations and infragravity waves in the surf and swash zone}} {{Remote
  sensing of the correlation between breakpoint oscillations and infragravity
  waves in the surf and swash zone}}.{\BBCQ}
\newblock
\APACjournalVolNumPages{Journal of Geophysical Research: Oceans}{}{}{1--17}.
\newblock
\begin{APACrefDOI} \doi{10.1002/2016JC012233.Received} \end{APACrefDOI}
\PrintBackRefs{\CurrentBib}

\bibitem [\protect \citeauthoryear {%
Freedman%
, Pisani%
\BCBL {}\ \BBA {} Purves%
}{%
Freedman%
\ \protect \BOthers {.}}{%
{\protect \APACyear {1998}}%
}]{%
Freedman1998}
\APACinsertmetastar {%
Freedman1998}%
\begin{APACrefauthors}%
Freedman, D.%
, Pisani, R.%
\BCBL {}\ \BBA {} Purves, R.%
\end{APACrefauthors}%
\unskip\
\newblock
\APACrefYear{1998}.
\newblock
\APACrefbtitle {{Statistics}} {{Statistics}}.
\newblock
\APACaddressPublisher{}{W.W. Norton}.
\newblock
\begin{APACrefURL} \url{https://books.google.com.au/books?id=wj3CQgAACAAJ}
  \end{APACrefURL}
\PrintBackRefs{\CurrentBib}

\bibitem [\protect \citeauthoryear {%
Galvin%
}{%
Galvin%
}{%
{\protect \APACyear {1968}}%
}]{%
Galvin1968}
\APACinsertmetastar {%
Galvin1968}%
\begin{APACrefauthors}%
Galvin, C\BPBI J.%
\end{APACrefauthors}%
\unskip\
\newblock
\APACrefYearMonthDay{1968}{}{}.
\newblock
{\BBOQ}\APACrefatitle {{Breaker type classification on three laboratory
  beaches}} {{Breaker type classification on three laboratory beaches}}.{\BBCQ}
\newblock
\APACjournalVolNumPages{Journal of Geophysical Research}{73}{12}{3651--3659}.
\newblock
\begin{APACrefDOI} \doi{10.1029/JB073i012p03651} \end{APACrefDOI}
\PrintBackRefs{\CurrentBib}

\bibitem [\protect \citeauthoryear {%
Garc{\'{i}}a-Medina%
, {\"{O}}zkan-Haller%
, Holman%
\BCBL {}\ \BBA {} Ruggiero%
}{%
Garc{\'{i}}a-Medina%
\ \protect \BOthers {.}}{%
{\protect \APACyear {2017}}%
}]{%
Garcia-Medina2017}
\APACinsertmetastar {%
Garcia-Medina2017}%
\begin{APACrefauthors}%
Garc{\'{i}}a-Medina, G.%
, {\"{O}}zkan-Haller, H\BPBI T.%
, Holman, R\BPBI A.%
\BCBL {}\ \BBA {} Ruggiero, P.%
\end{APACrefauthors}%
\unskip\
\newblock
\APACrefYearMonthDay{2017}{}{}.
\newblock
{\BBOQ}\APACrefatitle {{Large runup controls on a gently sloping dissipative
  beach}} {{Large runup controls on a gently sloping dissipative
  beach}}.{\BBCQ}
\newblock
\APACjournalVolNumPages{Journal of Geophysical Research: Oceans}{}{1}{1--20}.
\newblock
\begin{APACrefDOI} \doi{10.1002/2016JC012336.Received} \end{APACrefDOI}
\PrintBackRefs{\CurrentBib}

\bibitem [\protect \citeauthoryear {%
Gourlay%
}{%
Gourlay%
}{%
{\protect \APACyear {1968}}%
}]{%
Gourlay1968}
\APACinsertmetastar {%
Gourlay1968}%
\begin{APACrefauthors}%
Gourlay, M\BPBI R.%
\end{APACrefauthors}%
\unskip\
\newblock
\APACrefYearMonthDay{1968}{}{}.
\newblock
\APACrefbtitle {{Beach and dune erosion due to storms}.} {{Beach and dune
  erosion due to storms}.}
\PrintBackRefs{\CurrentBib}

\bibitem [\protect \citeauthoryear {%
Grinsted%
, Moore%
\BCBL {}\ \BBA {} Jevrejeva%
}{%
Grinsted%
\ \protect \BOthers {.}}{%
{\protect \APACyear {2004}}%
}]{%
Grinsted2004}
\APACinsertmetastar {%
Grinsted2004}%
\begin{APACrefauthors}%
Grinsted, A.%
, Moore, J\BPBI C.%
\BCBL {}\ \BBA {} Jevrejeva, S.%
\end{APACrefauthors}%
\unskip\
\newblock
\APACrefYearMonthDay{2004}{}{}.
\newblock
{\BBOQ}\APACrefatitle {{Application of the cross wavelet transform and wavelet
  coherence to geophysical time series}} {{Application of the cross wavelet
  transform and wavelet coherence to geophysical time series}}.{\BBCQ}
\newblock
\APACjournalVolNumPages{Nonlinear Processes in Geophysics}{11}{5/6}{561--566}.
\newblock
\begin{APACrefDOI} \doi{10.5194/npg-11-561-2004} \end{APACrefDOI}
\PrintBackRefs{\CurrentBib}

\bibitem [\protect \citeauthoryear {%
Guedes%
, Bryan%
\BCBL {}\ \BBA {} Coco%
}{%
Guedes%
\ \protect \BOthers {.}}{%
{\protect \APACyear {2013}}%
}]{%
Guedes2013}
\APACinsertmetastar {%
Guedes2013}%
\begin{APACrefauthors}%
Guedes, R\BPBI M\BPBI C.%
, Bryan, K\BPBI R.%
\BCBL {}\ \BBA {} Coco, G.%
\end{APACrefauthors}%
\unskip\
\newblock
\APACrefYearMonthDay{2013}{}{}.
\newblock
{\BBOQ}\APACrefatitle {{Observations of wave energy fluxes and swash motions on
  a low-sloping, dissipative beach}} {{Observations of wave energy fluxes and
  swash motions on a low-sloping, dissipative beach}}.{\BBCQ}
\newblock
\APACjournalVolNumPages{Journal of Geophysical Research:
  Oceans}{118}{7}{3651--3669}.
\newblock
\begin{APACrefDOI} \doi{10.1002/jgrc.20267} \end{APACrefDOI}
\PrintBackRefs{\CurrentBib}

\bibitem [\protect \citeauthoryear {%
Guza%
\ \BBA {} Thornton%
}{%
Guza%
\ \BBA {} Thornton%
}{%
{\protect \APACyear {1982}}%
}]{%
Guza1982}
\APACinsertmetastar {%
Guza1982}%
\begin{APACrefauthors}%
Guza, R\BPBI T.%
\BCBT {}\ \BBA {} Thornton, E\BPBI B.%
\end{APACrefauthors}%
\unskip\
\newblock
\APACrefYearMonthDay{1982}{}{}.
\newblock
{\BBOQ}\APACrefatitle {{Swash Oscillations on a Natural Beach}} {{Swash
  Oscillations on a Natural Beach}}.{\BBCQ}
\newblock
\APACjournalVolNumPages{Journal of Geophysical Research}{87}{1}{483--491}.
\PrintBackRefs{\CurrentBib}

\bibitem [\protect \citeauthoryear {%
Holland%
\ \protect \BOthers {.}}{%
Holland%
\ \protect \BOthers {.}}{%
{\protect \APACyear {1997}}%
}]{%
Holland1997}
\APACinsertmetastar {%
Holland1997}%
\begin{APACrefauthors}%
Holland, K\BPBI T.%
, Holman, R\BPBI A.%
, Lippmann, T\BPBI C.%
, Stanley, J.%
, Member, A.%
\BCBL {}\ \BBA {} Plant, N.%
\end{APACrefauthors}%
\unskip\
\newblock
\APACrefYearMonthDay{1997}{}{}.
\newblock
{\BBOQ}\APACrefatitle {{Practical Use of Video Imagery in Nearshore
  Oceanographic Field Studies}} {{Practical Use of Video Imagery in Nearshore
  Oceanographic Field Studies}}.{\BBCQ}
\newblock
\APACjournalVolNumPages{IEEE Journal of Oceanic Engineering}{22}{1}{81--92}.
\PrintBackRefs{\CurrentBib}

\bibitem [\protect \citeauthoryear {%
Holland%
, Raubenheimer%
, Guza%
\BCBL {}\ \BBA {} Holman%
}{%
Holland%
\ \protect \BOthers {.}}{%
{\protect \APACyear {1995}}%
}]{%
Holland1995}
\APACinsertmetastar {%
Holland1995}%
\begin{APACrefauthors}%
Holland, K\BPBI T.%
, Raubenheimer, B.%
, Guza, R\BPBI T.%
\BCBL {}\ \BBA {} Holman, R\BPBI A.%
\end{APACrefauthors}%
\unskip\
\newblock
\APACrefYearMonthDay{1995}{}{}.
\newblock
{\BBOQ}\APACrefatitle {{Runup kinematics on a natural beach}} {{Runup
  kinematics on a natural beach}}.{\BBCQ}
\newblock
\APACjournalVolNumPages{Journal of Geophysical Research}{100}{C3}{4985}.
\newblock
\begin{APACrefDOI} \doi{10.1029/94JC02664} \end{APACrefDOI}
\PrintBackRefs{\CurrentBib}

\bibitem [\protect \citeauthoryear {%
Holman%
}{%
Holman%
}{%
{\protect \APACyear {1986}}%
}]{%
Holman1986}
\APACinsertmetastar {%
Holman1986}%
\begin{APACrefauthors}%
Holman, R\BPBI A.%
\end{APACrefauthors}%
\unskip\
\newblock
\APACrefYearMonthDay{1986}{}{}.
\newblock
{\BBOQ}\APACrefatitle {{Extreme value statistics for wave run-up on a natural
  beach}} {{Extreme value statistics for wave run-up on a natural
  beach}}.{\BBCQ}
\newblock
\APACjournalVolNumPages{Coastal Engineering}{9}{6}{527--544}.
\newblock
\begin{APACrefDOI} \doi{10.1016/0378-3839(86)90002-5} \end{APACrefDOI}
\PrintBackRefs{\CurrentBib}

\bibitem [\protect \citeauthoryear {%
Holthuijsen%
}{%
Holthuijsen%
}{%
{\protect \APACyear {2010}}%
}]{%
holthuijsen2010}
\APACinsertmetastar {%
holthuijsen2010}%
\begin{APACrefauthors}%
Holthuijsen, L\BPBI H.%
\end{APACrefauthors}%
\unskip\
\newblock
\APACrefYear{2010}.
\newblock
\APACrefbtitle {{Waves in Oceanic and Coastal Waters}} {{Waves in Oceanic and
  Coastal Waters}}.
\newblock
\APACaddressPublisher{}{Cambridge University Press}.
\newblock
\begin{APACrefURL} \url{https://books.google.com.au/books?id=7tFUL2blHdoC}
  \end{APACrefURL}
\PrintBackRefs{\CurrentBib}

\bibitem [\protect \citeauthoryear {%
Hoonhout%
, Radermacher%
, Baart%
\BCBL {}\ \BBA {} Maaten%
}{%
Hoonhout%
\ \protect \BOthers {.}}{%
{\protect \APACyear {2015}}%
}]{%
Hoonhout2015}
\APACinsertmetastar {%
Hoonhout2015}%
\begin{APACrefauthors}%
Hoonhout, B\BPBI M.%
, Radermacher, M.%
, Baart, F.%
\BCBL {}\ \BBA {} Maaten, L\BPBI J\BPBI P\BPBI V\BPBI D.%
\end{APACrefauthors}%
\unskip\
\newblock
\APACrefYearMonthDay{2015}{}{}.
\newblock
{\BBOQ}\APACrefatitle {{An automated method for semantic classification of
  regions in coastal images}} {{An automated method for semantic classification
  of regions in coastal images}}.{\BBCQ}
\newblock
\APACjournalVolNumPages{Coastal Engineering}{105}{}{1--12}.
\newblock
\begin{APACrefURL} \url{http://dx.doi.org/10.1016/j.coastaleng.2015.07.010}
  \end{APACrefURL}
\newblock
\begin{APACrefDOI} \doi{10.1016/j.coastaleng.2015.07.010} \end{APACrefDOI}
\PrintBackRefs{\CurrentBib}

\bibitem [\protect \citeauthoryear {%
Hughes%
\ \BBA {} Moseley%
}{%
Hughes%
\ \BBA {} Moseley%
}{%
{\protect \APACyear {2007}}%
}]{%
Hughes2007}
\APACinsertmetastar {%
Hughes2007}%
\begin{APACrefauthors}%
Hughes, M\BPBI G.%
\BCBT {}\ \BBA {} Moseley, A\BPBI S.%
\end{APACrefauthors}%
\unskip\
\newblock
\APACrefYearMonthDay{2007}{}{}.
\newblock
{\BBOQ}\APACrefatitle {{Hydrokinematic regions within the swash zone}}
  {{Hydrokinematic regions within the swash zone}}.{\BBCQ}
\newblock
\APACjournalVolNumPages{Continental Shelf Research}{27}{15}{2000--2013}.
\newblock
\begin{APACrefDOI} \doi{10.1016/j.csr.2007.04.005} \end{APACrefDOI}
\PrintBackRefs{\CurrentBib}

\bibitem [\protect \citeauthoryear {%
Huntley%
}{%
Huntley%
}{%
{\protect \APACyear {1976}}%
}]{%
Huntley1976}
\APACinsertmetastar {%
Huntley1976}%
\begin{APACrefauthors}%
Huntley, D\BPBI a.%
\end{APACrefauthors}%
\unskip\
\newblock
\APACrefYearMonthDay{1976}{}{}.
\newblock
{\BBOQ}\APACrefatitle {{Long-period waves on a natural beach}} {{Long-period
  waves on a natural beach}}.{\BBCQ}
\newblock
\APACjournalVolNumPages{Journal of Geophysical Research}{81}{36}{6441}.
\newblock
\begin{APACrefDOI} \doi{10.1029/JC081i036p06441} \end{APACrefDOI}
\PrintBackRefs{\CurrentBib}

\bibitem [\protect \citeauthoryear {%
Iribarren%
\ \BBA {} Nogales%
}{%
Iribarren%
\ \BBA {} Nogales%
}{%
{\protect \APACyear {1949}}%
}]{%
Iribarren1949}
\APACinsertmetastar {%
Iribarren1949}%
\begin{APACrefauthors}%
Iribarren, C\BPBI R.%
\BCBT {}\ \BBA {} Nogales, C\BPBI M.%
\end{APACrefauthors}%
\unskip\
\newblock
\APACrefYearMonthDay{1949}{}{}.
\newblock
\APACrefbtitle {{Protection des ports}.} {{Protection des ports}.}
\newblock
\begin{APACrefURL}
  \url{http://cedb.asce.org/cgi/WWWdisplay.cgi?9401349{\%}5Cnhttp://repository.tudelft.nl/view/hydro/uuid:7ab718ff-a74d-4141-8c3f-413044c751c4/}
  \end{APACrefURL}
\PrintBackRefs{\CurrentBib}

\bibitem [\protect \citeauthoryear {%
Iske%
}{%
Iske%
}{%
{\protect \APACyear {2004}}%
}]{%
Iske2004}
\APACinsertmetastar {%
Iske2004}%
\begin{APACrefauthors}%
Iske, A.%
\end{APACrefauthors}%
\unskip\
\newblock
\APACrefYearMonthDay{2004}{}{}.
\newblock
\APACrefbtitle {{Radial Basis Functions}} {{Radial Basis Functions}}\
  \APACbVolEdTR{}{\BTR{}}.
\PrintBackRefs{\CurrentBib}

\bibitem [\protect \citeauthoryear {%
Jones%
, Oliphant%
, Peterson%
\BCBL {}\ \BBA {} Others%
}{%
Jones%
\ \protect \BOthers {.}}{%
{\protect \APACyear {2001}}%
}]{%
Jones2001}
\APACinsertmetastar {%
Jones2001}%
\begin{APACrefauthors}%
Jones, E.%
, Oliphant, T.%
, Peterson, P.%
\BCBL {}\ \BBA {} Others.%
\end{APACrefauthors}%
\unskip\
\newblock
\APACrefYearMonthDay{2001}{}{}.
\newblock
\APACrefbtitle {{{\{}SciPy{\}}: Open source scientific tools for
  {\{}Python{\}}}.} {{{\{}SciPy{\}}: Open source scientific tools for
  {\{}Python{\}}}.}
\newblock
\begin{APACrefURL} \url{http://www.scipy.org/} \end{APACrefURL}
\PrintBackRefs{\CurrentBib}

\bibitem [\protect \citeauthoryear {%
Komar%
}{%
Komar%
}{%
{\protect \APACyear {1976}}%
}]{%
komar1976beach}
\APACinsertmetastar {%
komar1976beach}%
\begin{APACrefauthors}%
Komar, P\BPBI D.%
\end{APACrefauthors}%
\unskip\
\newblock
\APACrefYear{1976}.
\newblock
\APACrefbtitle {{Beach Processes and Sedimentation}} {{Beach Processes and
  Sedimentation}}.
\newblock
\APACaddressPublisher{}{Prentice-Hall}.
\newblock
\begin{APACrefURL} \url{https://books.google.com.au/books?id=-vNOAAAAMAAJ}
  \end{APACrefURL}
\PrintBackRefs{\CurrentBib}

\bibitem [\protect \citeauthoryear {%
Longuet-Higgins%
\ \BBA {} Stewart%
}{%
Longuet-Higgins%
\ \BBA {} Stewart%
}{%
{\protect \APACyear {1964}}%
}]{%
LonguetHiggins1964}
\APACinsertmetastar {%
LonguetHiggins1964}%
\begin{APACrefauthors}%
Longuet-Higgins, M\BPBI S.%
\BCBT {}\ \BBA {} Stewart, R\BPBI W.%
\end{APACrefauthors}%
\unskip\
\newblock
\APACrefYearMonthDay{1964}{}{}.
\newblock
{\BBOQ}\APACrefatitle {{Radiation stresses in water waves; a physical
  discussion, with applications}} {{Radiation stresses in water waves; a
  physical discussion, with applications}}.{\BBCQ}
\newblock
\APACjournalVolNumPages{Deep Sea Research and Oceanographic
  Abstracts}{11}{4}{529--562}.
\newblock
\begin{APACrefURL}
  \url{http://www.sciencedirect.com/science/article/pii/0011747164900014}
  \end{APACrefURL}
\newblock
\begin{APACrefDOI} \doi{10.1016/0011-7471(64)90001-4} \end{APACrefDOI}
\PrintBackRefs{\CurrentBib}

\bibitem [\protect \citeauthoryear {%
MacMahan%
, Reniers%
\BCBL {}\ \BBA {} Thornton%
}{%
MacMahan%
\ \protect \BOthers {.}}{%
{\protect \APACyear {2010}}%
}]{%
MacMahan2010}
\APACinsertmetastar {%
MacMahan2010}%
\begin{APACrefauthors}%
MacMahan, J\BPBI H.%
, Reniers, A\BPBI J\BPBI H\BPBI M.%
\BCBL {}\ \BBA {} Thornton, E\BPBI B.%
\end{APACrefauthors}%
\unskip\
\newblock
\APACrefYearMonthDay{2010}{}{}.
\newblock
{\BBOQ}\APACrefatitle {{Vortical surf zone velocity fluctuations with 0(10) min
  period}} {{Vortical surf zone velocity fluctuations with 0(10) min
  period}}.{\BBCQ}
\newblock
\APACjournalVolNumPages{Journal of Geophysical Research:
  Oceans}{115}{6}{1--18}.
\newblock
\begin{APACrefDOI} \doi{10.1029/2009JC005383} \end{APACrefDOI}
\PrintBackRefs{\CurrentBib}

\bibitem [\protect \citeauthoryear {%
Masselink%
\ \BBA {} Puleo%
}{%
Masselink%
\ \BBA {} Puleo%
}{%
{\protect \APACyear {2006}}%
}]{%
Masselink2006}
\APACinsertmetastar {%
Masselink2006}%
\begin{APACrefauthors}%
Masselink, G.%
\BCBT {}\ \BBA {} Puleo, J\BPBI A.%
\end{APACrefauthors}%
\unskip\
\newblock
\APACrefYearMonthDay{2006}{}{}.
\newblock
{\BBOQ}\APACrefatitle {{Swash-zone morphodynamics}} {{Swash-zone
  morphodynamics}}.{\BBCQ}
\newblock
\APACjournalVolNumPages{Continental Shelf Research}{26}{5}{661--680}.
\newblock
\begin{APACrefDOI} \doi{10.1016/j.csr.2006.01.015} \end{APACrefDOI}
\PrintBackRefs{\CurrentBib}

\bibitem [\protect \citeauthoryear {%
Moura%
\ \BBA {} Baldock%
}{%
Moura%
\ \BBA {} Baldock%
}{%
{\protect \APACyear {2018}}%
}]{%
Moura2018}
\APACinsertmetastar {%
Moura2018}%
\begin{APACrefauthors}%
Moura, T.%
\BCBT {}\ \BBA {} Baldock, T\BPBI E.%
\end{APACrefauthors}%
\unskip\
\newblock
\APACrefYearMonthDay{2018}{}{}.
\newblock
{\BBOQ}\APACrefatitle {{New Evidence of Breakpoint Forced Long Waves:
  Laboratory, Numerical, and Field Observations}} {{New Evidence of Breakpoint
  Forced Long Waves: Laboratory, Numerical, and Field Observations}}.{\BBCQ}
\newblock
\APACjournalVolNumPages{Journal of Geophysical Research:
  Oceans}{123}{4}{2716--2730}.
\newblock
\begin{APACrefDOI} \doi{10.1002/2017JC013621} \end{APACrefDOI}
\PrintBackRefs{\CurrentBib}

\bibitem [\protect \citeauthoryear {%
Nam%
, Larson%
, Hanson%
\BCBL {}\ \BBA {} Hoan%
}{%
Nam%
\ \protect \BOthers {.}}{%
{\protect \APACyear {2009}}%
}]{%
Nam2009}
\APACinsertmetastar {%
Nam2009}%
\begin{APACrefauthors}%
Nam, P\BPBI T.%
, Larson, M.%
, Hanson, H.%
\BCBL {}\ \BBA {} Hoan, L\BPBI X.%
\end{APACrefauthors}%
\unskip\
\newblock
\APACrefYearMonthDay{2009}{}{}.
\newblock
{\BBOQ}\APACrefatitle {{A numerical model of nearshore waves, currents, and
  sediment transport}} {{A numerical model of nearshore waves, currents, and
  sediment transport}}.{\BBCQ}
\newblock
\APACjournalVolNumPages{Coastal Engineering}{56}{11-12}{1084--1096}.
\newblock
\begin{APACrefURL} \url{http://dx.doi.org/10.1016/j.coastaleng.2009.06.007}
  \end{APACrefURL}
\newblock
\begin{APACrefDOI} \doi{10.1016/j.coastaleng.2009.06.007} \end{APACrefDOI}
\PrintBackRefs{\CurrentBib}

\bibitem [\protect \citeauthoryear {%
Nielsen%
\ \BBA {} Hanslow%
}{%
Nielsen%
\ \BBA {} Hanslow%
}{%
{\protect \APACyear {1991}}%
}]{%
Nielsen1991}
\APACinsertmetastar {%
Nielsen1991}%
\begin{APACrefauthors}%
Nielsen, P.%
\BCBT {}\ \BBA {} Hanslow, D\BPBI J.%
\end{APACrefauthors}%
\unskip\
\newblock
\APACrefYearMonthDay{1991}{}{}.
\newblock
{\BBOQ}\APACrefatitle {{of Coastal Fall Wave Runup Distributions on Natural
  Beaches}} {{of Coastal Fall Wave Runup Distributions on Natural
  Beaches}}.{\BBCQ}
\newblock
\APACjournalVolNumPages{Journal of Coastal Research}{7}{4}{1139--1152}.
\PrintBackRefs{\CurrentBib}

\bibitem [\protect \citeauthoryear {%
Plant%
\ \BBA {} Stockdon%
}{%
Plant%
\ \BBA {} Stockdon%
}{%
{\protect \APACyear {2015}}%
}]{%
Plant2015}
\APACinsertmetastar {%
Plant2015}%
\begin{APACrefauthors}%
Plant, N\BPBI G.%
\BCBT {}\ \BBA {} Stockdon, H\BPBI F.%
\end{APACrefauthors}%
\unskip\
\newblock
\APACrefYearMonthDay{2015}{}{}.
\newblock
{\BBOQ}\APACrefatitle {{How well can wave runup be predicted? Comment on
  Laudier et al. (2011) and Stockdon et al. (2006)}} {{How well can wave runup
  be predicted? Comment on Laudier et al. (2011) and Stockdon et al.
  (2006)}}.{\BBCQ}
\newblock
\APACjournalVolNumPages{Coastal Engineering}{102}{}{44--48}.
\newblock
\begin{APACrefURL} \url{http://dx.doi.org/10.1016/j.coastaleng.2015.05.001}
  \end{APACrefURL}
\newblock
\begin{APACrefDOI} \doi{10.1016/j.coastaleng.2015.05.001} \end{APACrefDOI}
\PrintBackRefs{\CurrentBib}

\bibitem [\protect \citeauthoryear {%
Power%
, Hughes%
\BCBL {}\ \BBA {} Baldock%
}{%
Power%
\ \protect \BOthers {.}}{%
{\protect \APACyear {2015}}%
}]{%
Power2015}
\APACinsertmetastar {%
Power2015}%
\begin{APACrefauthors}%
Power, H\BPBI E.%
, Hughes, M\BPBI G.%
\BCBL {}\ \BBA {} Baldock, T\BPBI E.%
\end{APACrefauthors}%
\unskip\
\newblock
\APACrefYearMonthDay{2015}{}{}.
\newblock
{\BBOQ}\APACrefatitle {{A novel method for tracking individual waves in the
  surf zone}} {{A novel method for tracking individual waves in the surf
  zone}}.{\BBCQ}
\newblock
\APACjournalVolNumPages{Coastal Engineering}{98}{}{26--30}.
\newblock
\begin{APACrefURL} \url{http://dx.doi.org/10.1016/j.coastaleng.2015.01.006}
  \end{APACrefURL}
\newblock
\begin{APACrefDOI} \doi{10.1016/j.coastaleng.2015.01.006} \end{APACrefDOI}
\PrintBackRefs{\CurrentBib}

\bibitem [\protect \citeauthoryear {%
Puleo%
\ \BBA {} Holland%
}{%
Puleo%
\ \BBA {} Holland%
}{%
{\protect \APACyear {2001}}%
}]{%
Puleo2001}
\APACinsertmetastar {%
Puleo2001}%
\begin{APACrefauthors}%
Puleo, J\BPBI A.%
\BCBT {}\ \BBA {} Holland, K\BPBI T.%
\end{APACrefauthors}%
\unskip\
\newblock
\APACrefYearMonthDay{2001}{}{}.
\newblock
{\BBOQ}\APACrefatitle {{Estimating swash zone friction coefficients on a sandy
  beach}} {{Estimating swash zone friction coefficients on a sandy
  beach}}.{\BBCQ}
\newblock
\APACjournalVolNumPages{Coastal Engineering}{43}{1}{25--40}.
\newblock
\begin{APACrefDOI} \doi{10.1016/S0378-3839(01)00004-7} \end{APACrefDOI}
\PrintBackRefs{\CurrentBib}

\bibitem [\protect \citeauthoryear {%
Russell%
}{%
Russell%
}{%
{\protect \APACyear {1993}}%
}]{%
Russell1993}
\APACinsertmetastar {%
Russell1993}%
\begin{APACrefauthors}%
Russell, P\BPBI E.%
\end{APACrefauthors}%
\unskip\
\newblock
\APACrefYearMonthDay{1993}{}{}.
\newblock
{\BBOQ}\APACrefatitle {{Mechanisms for beach erosion during storms}}
  {{Mechanisms for beach erosion during storms}}.{\BBCQ}
\newblock
\APACjournalVolNumPages{Continental Shelf Research}{13}{11}{1243--1265}.
\newblock
\begin{APACrefDOI} \doi{10.1016/0278-4343(93)90051-X} \end{APACrefDOI}
\PrintBackRefs{\CurrentBib}

\bibitem [\protect \citeauthoryear {%
S{\'{e}}n{\'{e}}chal%
, Dupuis%
, Bonneton%
, Howa%
\BCBL {}\ \BBA {} Pedreros%
}{%
S{\'{e}}n{\'{e}}chal%
\ \protect \BOthers {.}}{%
{\protect \APACyear {2001}}%
}]{%
Senechal2001}
\APACinsertmetastar {%
Senechal2001}%
\begin{APACrefauthors}%
S{\'{e}}n{\'{e}}chal, N.%
, Dupuis, H.%
, Bonneton, P.%
, Howa, H.%
\BCBL {}\ \BBA {} Pedreros, R.%
\end{APACrefauthors}%
\unskip\
\newblock
\APACrefYearMonthDay{2001}{}{}.
\newblock
{\BBOQ}\APACrefatitle {{Observation of irregular wave transformation in the
  surf zone over a gently sloping sandy beach on the French Atlantic
  coastline}} {{Observation of irregular wave transformation in the surf zone
  over a gently sloping sandy beach on the French Atlantic coastline}}.{\BBCQ}
\newblock
\APACjournalVolNumPages{Oceanologica Acta}{24}{6}{545--556}.
\PrintBackRefs{\CurrentBib}

\bibitem [\protect \citeauthoryear {%
Shen%
\ \BBA {} Meyer%
}{%
Shen%
\ \BBA {} Meyer%
}{%
{\protect \APACyear {1963}}%
}]{%
shen1963}
\APACinsertmetastar {%
shen1963}%
\begin{APACrefauthors}%
Shen, M\BPBI C.%
\BCBT {}\ \BBA {} Meyer, R\BPBI E.%
\end{APACrefauthors}%
\unskip\
\newblock
\APACrefYearMonthDay{1963}{}{}.
\newblock
{\BBOQ}\APACrefatitle {{Climb of a bore on a beach. Part 3. Run-up}} {{Climb of
  a bore on a beach. Part 3. Run-up}}.{\BBCQ}
\newblock
\APACjournalVolNumPages{Journal of Fluid Mechanics}{14}{02}{305--318}.
\newblock
\begin{APACrefURL}
  \url{http://journals.cambridge.org/production/action/cjoGetFulltext?fulltextid=369325{\%}5Cnhttp://www.journals.cambridge.org/abstract{\_}S0022112062001251}
  \end{APACrefURL}
\newblock
\begin{APACrefDOI} \doi{10.1017/S0022112062001251} \end{APACrefDOI}
\PrintBackRefs{\CurrentBib}

\bibitem [\protect \citeauthoryear {%
Short%
}{%
Short%
}{%
{\protect \APACyear {1999}}%
}]{%
short1999beaches}
\APACinsertmetastar {%
short1999beaches}%
\begin{APACrefauthors}%
Short, A\BPBI D.%
\end{APACrefauthors}%
\unskip\
\newblock
\APACrefYear{1999}.
\newblock
\APACrefbtitle {{Beaches of the Queensland Coast, Cooktown to Coolangatta: A
  Guide to Their Nature, Characteristics, Surf and Safety}} {{Beaches of the
  Queensland Coast, Cooktown to Coolangatta: A Guide to Their Nature,
  Characteristics, Surf and Safety}}.
\newblock
\APACaddressPublisher{}{Sydney University Press}.
\newblock
\begin{APACrefURL} \url{https://books.google.com.au/books?id=qEs92AQ5gfYC}
  \end{APACrefURL}
\PrintBackRefs{\CurrentBib}

\bibitem [\protect \citeauthoryear {%
Short%
}{%
Short%
}{%
{\protect \APACyear {2007}}%
}]{%
short2007beaches}
\APACinsertmetastar {%
short2007beaches}%
\begin{APACrefauthors}%
Short, A\BPBI D.%
\end{APACrefauthors}%
\unskip\
\newblock
\APACrefYear{2007}.
\newblock
\APACrefbtitle {{Beaches of the New South Wales Coast: A Guide to Their Nature,
  Characteristics, Surf and Safety}} {{Beaches of the New South Wales Coast: A
  Guide to Their Nature, Characteristics, Surf and Safety}}.
\newblock
\APACaddressPublisher{}{Sydney University Press}.
\newblock
\begin{APACrefURL} \url{https://books.google.com.au/books?id=eeopVw9kedgC}
  \end{APACrefURL}
\PrintBackRefs{\CurrentBib}

\bibitem [\protect \citeauthoryear {%
Stockdon%
, Holman%
, Howd%
\BCBL {}\ \BBA {} Sallenger%
}{%
Stockdon%
\ \protect \BOthers {.}}{%
{\protect \APACyear {2006}}%
}]{%
Stockdon2006}
\APACinsertmetastar {%
Stockdon2006}%
\begin{APACrefauthors}%
Stockdon, H\BPBI F.%
, Holman, R\BPBI A.%
, Howd, P\BPBI A.%
\BCBL {}\ \BBA {} Sallenger, A\BPBI H.%
\end{APACrefauthors}%
\unskip\
\newblock
\APACrefYearMonthDay{2006}{}{}.
\newblock
{\BBOQ}\APACrefatitle {{Empirical parameterization of setup, swash, and runup}}
  {{Empirical parameterization of setup, swash, and runup}}.{\BBCQ}
\newblock
\APACjournalVolNumPages{Coastal Engineering}{53}{7}{573--588}.
\newblock
\begin{APACrefDOI} \doi{10.1016/j.coastaleng.2005.12.005} \end{APACrefDOI}
\PrintBackRefs{\CurrentBib}

\bibitem [\protect \citeauthoryear {%
Stringari%
, Harris%
\BCBL {}\ \BBA {} Power%
}{%
Stringari%
\ \protect \BOthers {.}}{%
{\protect \APACyear {2019}}%
}]{%
Stringari2019}
\APACinsertmetastar {%
Stringari2019}%
\begin{APACrefauthors}%
Stringari, C\BPBI E.%
, Harris, D\BPBI L.%
\BCBL {}\ \BBA {} Power, H\BPBI E.%
\end{APACrefauthors}%
\unskip\
\newblock
\APACrefYearMonthDay{2019}{}{}.
\newblock
{\BBOQ}\APACrefatitle {{A Novel Machine Learning Algorithm for Tracking
  Remotely Sensed Waves in the Surf Zone}} {{A Novel Machine Learning Algorithm
  for Tracking Remotely Sensed Waves in the Surf Zone}}.{\BBCQ}
\newblock
\APACjournalVolNumPages{Coastal Engineering}{147}{}{149--158}.
\PrintBackRefs{\CurrentBib}

\bibitem [\protect \citeauthoryear {%
Svendsen%
}{%
Svendsen%
}{%
{\protect \APACyear {2006}}%
}]{%
svendsen2006}
\APACinsertmetastar {%
svendsen2006}%
\begin{APACrefauthors}%
Svendsen, I\BPBI A.%
\end{APACrefauthors}%
\unskip\
\newblock
\APACrefYear{2006}.
\newblock
\APACrefbtitle {{Introduction to Nearshore Hydrodynamics}} {{Introduction to
  Nearshore Hydrodynamics}}.
\newblock
\APACaddressPublisher{}{World Scientific}.
\PrintBackRefs{\CurrentBib}

\bibitem [\protect \citeauthoryear {%
Svendsen%
, Qin%
\BCBL {}\ \BBA {} Ebersole%
}{%
Svendsen%
\ \protect \BOthers {.}}{%
{\protect \APACyear {2003}}%
}]{%
Svendsen2003}
\APACinsertmetastar {%
Svendsen2003}%
\begin{APACrefauthors}%
Svendsen, I\BPBI A.%
, Qin, W.%
\BCBL {}\ \BBA {} Ebersole, B\BPBI A.%
\end{APACrefauthors}%
\unskip\
\newblock
\APACrefYearMonthDay{2003}{}{}.
\newblock
{\BBOQ}\APACrefatitle {{Modelling waves and currents at the LSTF and other
  laboratory facilities}} {{Modelling waves and currents at the LSTF and other
  laboratory facilities}}.{\BBCQ}
\newblock
\APACjournalVolNumPages{Coastal Engineering}{50}{1-2}{19--45}.
\newblock
\begin{APACrefDOI} \doi{10.1016/S0378-3839(03)00077-2} \end{APACrefDOI}
\PrintBackRefs{\CurrentBib}

\bibitem [\protect \citeauthoryear {%
Symonds%
, Huntley%
\BCBL {}\ \BBA {} Bowen%
}{%
Symonds%
\ \protect \BOthers {.}}{%
{\protect \APACyear {1982}}%
}]{%
Symonds1982}
\APACinsertmetastar {%
Symonds1982}%
\begin{APACrefauthors}%
Symonds, G.%
, Huntley, D\BPBI A.%
\BCBL {}\ \BBA {} Bowen, A\BPBI J.%
\end{APACrefauthors}%
\unskip\
\newblock
\APACrefYearMonthDay{1982}{}{}.
\newblock
{\BBOQ}\APACrefatitle {{Two-dimensional surf beat: Long wave generation by a
  time-varying breakpoint}} {{Two-dimensional surf beat: Long wave generation
  by a time-varying breakpoint}}.{\BBCQ}
\newblock
\APACjournalVolNumPages{Journal of Geophysical Research}{87}{1}{492--498}.
\PrintBackRefs{\CurrentBib}

\bibitem [\protect \citeauthoryear {%
Tissier%
, Bonneton%
, Michallet%
\BCBL {}\ \BBA {} Ruessink%
}{%
Tissier%
\ \protect \BOthers {.}}{%
{\protect \APACyear {2015}}%
}]{%
Tissier2015}
\APACinsertmetastar {%
Tissier2015}%
\begin{APACrefauthors}%
Tissier, M.%
, Bonneton, P.%
, Michallet, H.%
\BCBL {}\ \BBA {} Ruessink, B\BPBI G.%
\end{APACrefauthors}%
\unskip\
\newblock
\APACrefYearMonthDay{2015}{}{}.
\newblock
{\BBOQ}\APACrefatitle {{Infragravity-wave modulation of short-wave celerity in
  the surf zone}} {{Infragravity-wave modulation of short-wave celerity in the
  surf zone}}.{\BBCQ}
\newblock
\APACjournalVolNumPages{Journal of Geophysical Research:
  Oceans}{120}{10}{6799--6814}.
\newblock
\begin{APACrefDOI} \doi{10.1002/2015JC010708} \end{APACrefDOI}
\PrintBackRefs{\CurrentBib}

\bibitem [\protect \citeauthoryear {%
Torrence%
\ \BBA {} Compo%
}{%
Torrence%
\ \BBA {} Compo%
}{%
{\protect \APACyear {1998}}%
}]{%
Torrence1998}
\APACinsertmetastar {%
Torrence1998}%
\begin{APACrefauthors}%
Torrence, C.%
\BCBT {}\ \BBA {} Compo, G\BPBI P.%
\end{APACrefauthors}%
\unskip\
\newblock
\APACrefYearMonthDay{1998}{}{}.
\newblock
{\BBOQ}\APACrefatitle {{A Practical Guide to Wavelet Analysis}} {{A Practical
  Guide to Wavelet Analysis}}.{\BBCQ}
\newblock
\APACjournalVolNumPages{Bulletin of American Meteorological
  Analysis}{79}{1}{61--78}.
\newblock
\begin{APACrefURL}
  \url{http://paos.colorado.edu/research/wavelets/bams{\_}79{\_}01{\_}0061.pdf
  http://paos.colorado.edu/research/wavelets/bams{\%}7B{\_}{\%}7D79{\%}7B{\_}{\%}7D01{\%}7B{\_}{\%}7D0061.pdf}
  \end{APACrefURL}
\newblock
\begin{APACrefDOI} \doi{10.1175/1520-0477(1998)079<0061:APGTWA>2.0.CO;2}
  \end{APACrefDOI}
\PrintBackRefs{\CurrentBib}

\bibitem [\protect \citeauthoryear {%
van Dongeren%
\ \protect \BOthers {.}}{%
van Dongeren%
\ \protect \BOthers {.}}{%
{\protect \APACyear {2007}}%
}]{%
VanDongeren2007}
\APACinsertmetastar {%
VanDongeren2007}%
\begin{APACrefauthors}%
van Dongeren, A.%
, Battjes, J\BPBI A.%
, Janssen, T.%
, van Noorloos, J.%
, Steenhauer, K.%
, Steenbergen, G.%
\BCBL {}\ \BBA {} Reniers, A.%
\end{APACrefauthors}%
\unskip\
\newblock
\APACrefYearMonthDay{2007}{}{}.
\newblock
{\BBOQ}\APACrefatitle {{Shoaling and shoreline dissipation of low-frequency
  waves}} {{Shoaling and shoreline dissipation of low-frequency waves}}.{\BBCQ}
\newblock
\APACjournalVolNumPages{Journal of Geophysical Research:
  Oceans}{112}{2}{1--15}.
\newblock
\begin{APACrefDOI} \doi{10.1029/2006JC003701} \end{APACrefDOI}
\PrintBackRefs{\CurrentBib}

\bibitem [\protect \citeauthoryear {%
Wright%
\ \BBA {} Short%
}{%
Wright%
\ \BBA {} Short%
}{%
{\protect \APACyear {1984}}%
}]{%
Wright1984}
\APACinsertmetastar {%
Wright1984}%
\begin{APACrefauthors}%
Wright, L\BPBI D.%
\BCBT {}\ \BBA {} Short, A\BPBI D.%
\end{APACrefauthors}%
\unskip\
\newblock
\APACrefYearMonthDay{1984}{}{}.
\newblock
{\BBOQ}\APACrefatitle {{Morphodynamic variability of surf zones and beaches: A
  synthesis}} {{Morphodynamic variability of surf zones and beaches: A
  synthesis}}.{\BBCQ}
\newblock
\APACjournalVolNumPages{Marine Geology}{56}{1-4}{93--118}.
\newblock
\begin{APACrefDOI} \doi{10.1016/0025-3227(84)90008-2} \end{APACrefDOI}
\PrintBackRefs{\CurrentBib}

\end{thebibliography}

%
%
%
%
%

\end{document}